\documentclass[conference]{IEEEtran}

\IEEEoverridecommandlockouts
\usepackage{cite}
\usepackage{amsmath,amssymb,amsfonts}
\usepackage{algorithmic}
\usepackage{textcomp}
\usepackage{xcolor}
\def\BibTeX{{\rm B\kern-.05em{\sc i\kern-.025em b}\kern-.08em
    T\kern-.1667em\lower.7ex\hbox{E}\kern-.125emX}}

\usepackage{lipsum}
\usepackage{multirow, graphicx}
\ifCLASSOPTIONcompsoc
    \usepackage[caption=false, font=normalsize, labelfont=sf, textfont=sf]{subfig}
\else
\usepackage[caption=false, font=footnotesize]{subfig}
\fi

\usepackage{amsmath} 
\renewcommand{\vec}[1]{\boldsymbol{#1}}

\begin{document}\title{Market Manipulation of Bitcoin: Evidence from Mining the Mt. Gox Transaction Network
}




 \author{\IEEEauthorblockN{Weili Chen, Jun Wu, Zibin Zheng\IEEEauthorrefmark{1}, Chuan Chen\IEEEauthorrefmark{1}, and Yuren Zhou}\thanks{\IEEEauthorrefmark{1} Zibin Zheng and Chuan Chen are both corresponding authors.}
School of Data and Computer Science, Sun Yat-sen University, Guangzhou, China\\
National Engineering Research Center of Digital Life, Sun Yat-sen University, Guangzhou, 510006, China\\
\{chenwli9, wujun53\}@mail2.sysu.edu.cn, \{zhzibin, chenchuan, zhouyuren\}@mail.sysu.edu.cn \\
}

\maketitle

\begin{abstract}
The cryptocurrency market is a very huge market without effective supervision. It is of great importance for investors and regulators to recognize whether there are market manipulation and its manipulation patterns. This paper proposes an approach to mine the transaction networks of exchanges for answering this question. By taking the leaked transaction history of Mt. Gox Bitcoin exchange as a sample, we first divide the accounts into three categories according to its characteristic and then construct the transaction history into three graphs. Many observations and findings are obtained via analyzing the constructed graphs. To evaluate the influence of the accounts' transaction behavior on the Bitcoin exchange price, the graphs are reconstructed into series and reshaped as matrices. By using singular value decomposition (SVD) on the matrices, we identify many base networks which have a great correlation with the price fluctuation. When further analyzing the most important accounts in the base networks, plenty of market manipulation patterns are found. According to these findings, we conclude that there was serious market manipulation in Mt. Gox exchange and the cryptocurrency market must strengthen the supervision. 

\end{abstract}


\section{Introduction}
Bitcoin has become one of the hottest buzzwords among investors and researchers. It is the first and most famous decentralized digital currency\cite{nakamoto2008bitcoin}, which is secured by cryptography (thus, we call it cryptocurrency). Unlike fiat currencies which usually issued by financial institutions, there is no centralized organization or country controlling the issue and operation of Bitcoin. Furthermore, because of decentralization, users in the Bitcoin system are anonymous. The two characteristics (i.e., decentralization and anonymity) make Bitcoin attract a lot of users since its creation in 2009. It is estimated that there are more than 10 million users in the Bitcoin system \cite{burniske2017bitcoin}. 

Since the famous ``Bitcoin Pizza Day'' when a programmer bought two pizzas with 10,000 BTC on May 22, 2010, Bitcoin began to exchange with fiat currencies. Soon afterward, a Bitcoin exchange, Mt. Gox launched. By 2013 and before filing for bankruptcy protection in February 2014, Mt. Gox was the largest bitcoin intermediary and the world's leading Bitcoin exchange \cite{feder2018impact}. Nowadays, there are more than 1,700 cryptocurrencies inspired by Bitcoin and the daily transaction volume is over \$ 150 billion dollar according to coinmarketcap.com at the moment of writing this paper.    

\begin{figure}[htbp]
\centering
\includegraphics[width=.48\textwidth]{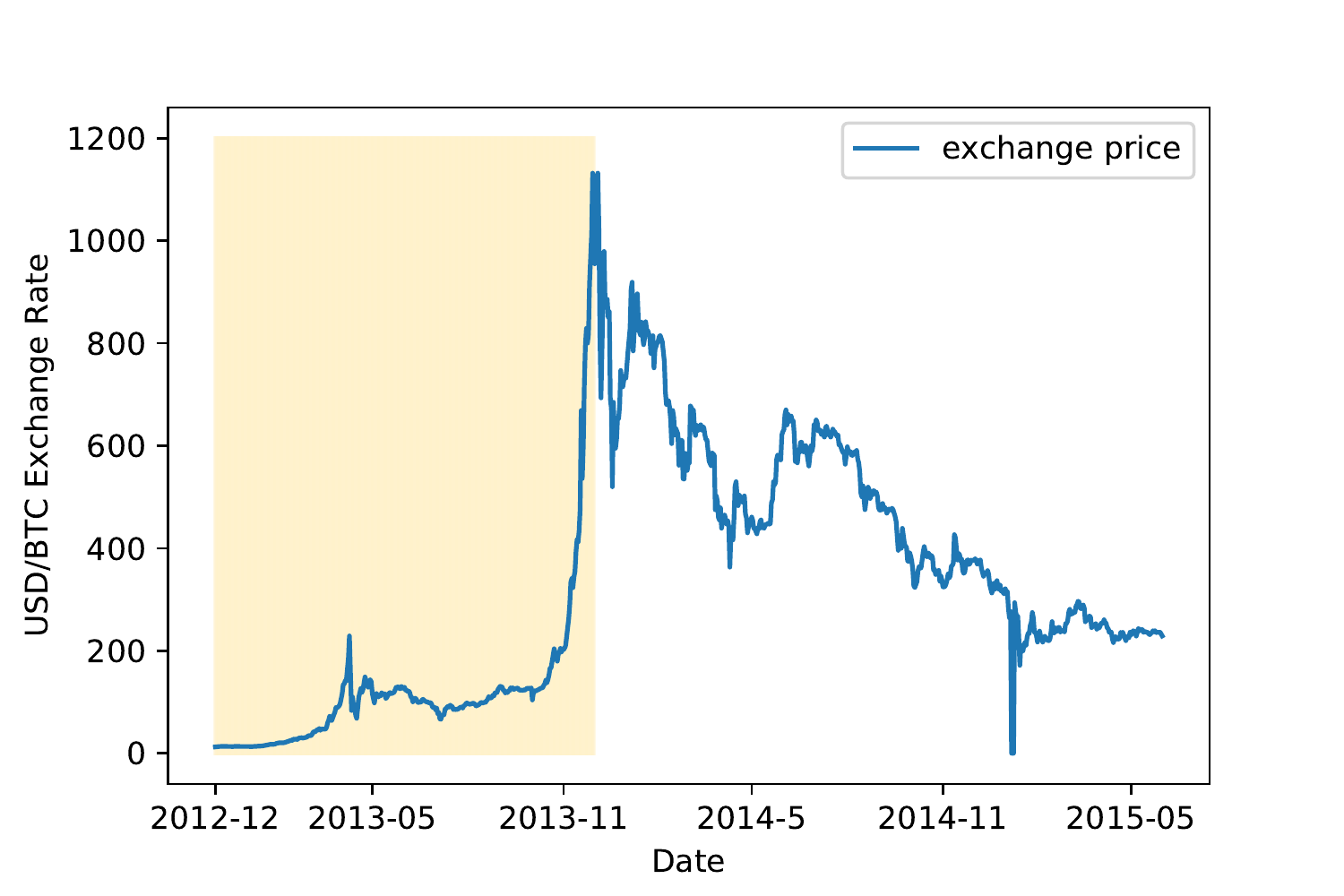}
\caption{Bitcoin-USD exchange price at Bitstamp exchange, with the period being studied shaded.}\label{fig_price}
\end{figure}

The huge fluctuation of the exchange price of cryptocurrency is an important reason to attract investors' participation. Figure \ref{fig_price} shows the Bitcoin price (i.e., the exchange rate between Bitcoin and USD dollar in this paper) from 2012/12 to 2015/6. During this period, the Bitcoin price rose sharply from about \$10/BTC to exceeding \$1,000/BTC and then fell back to below \$200/BTC. This extreme price fluctuation has also attracted a large number of researchers to find the determinant factors of the Bitcoin price. Four categories of factors are discussed, including 1) economic factors (e.g., the supply and demand of Bitcoin) \cite{buchholz2012bits}; 2) technical factors (e.g., hash rate and difficulty) \cite{kristoufek2015main}; 3) interest factors (through proxy variable such as Google trends) \cite{kristoufek2013bitcoin}; and 4) other financial assets (e.g., gold, stock). In addition, by using the principal component analysis method (analogous to SVD), the paper \cite{kondor2014inferring} indicates that the Bitcoin price has a strong correlation with the transactions on the blockchain ledger.  


However, these factors are discussed based on data outside the exchanges. Because of the lack of supervision, a nature conjecture is that the extreme fluctuation may be related to the market manipulation of the exchanges. This conjecture is hard to verify as it is very difficult to obtain the detailed trading data from the trading platform. Surprisingly, many transaction histories from April 2011 to November 2013 of the once famous Bitcoin exchange Mt. Gox were leaked in the form of CSV files. These data provide a perfect opportunity for answering the conjecture. 

To verify whether there is market manipulation and identify possible manipulation patterns is urgent and of great importance, as plenty of investors who are dreaming of getting rich overnight are attracted to the market. The answer to this question will help investors recognize the potential risks and help to regulate legislation. Based on the leaked data, a recent paper~\cite{gandal2018price} points out that the Mt. Gox exchange manipulated the Bitcoin price by building a regression model to identify the influence of the activities of some suspicious accounts on the price. We adopt a completely different method compared with it and obtain more results including fake volume, price manipulation, and manipulation patterns.

\begin{figure}[htbp]

\centering
 \includegraphics[width=.47\textwidth]{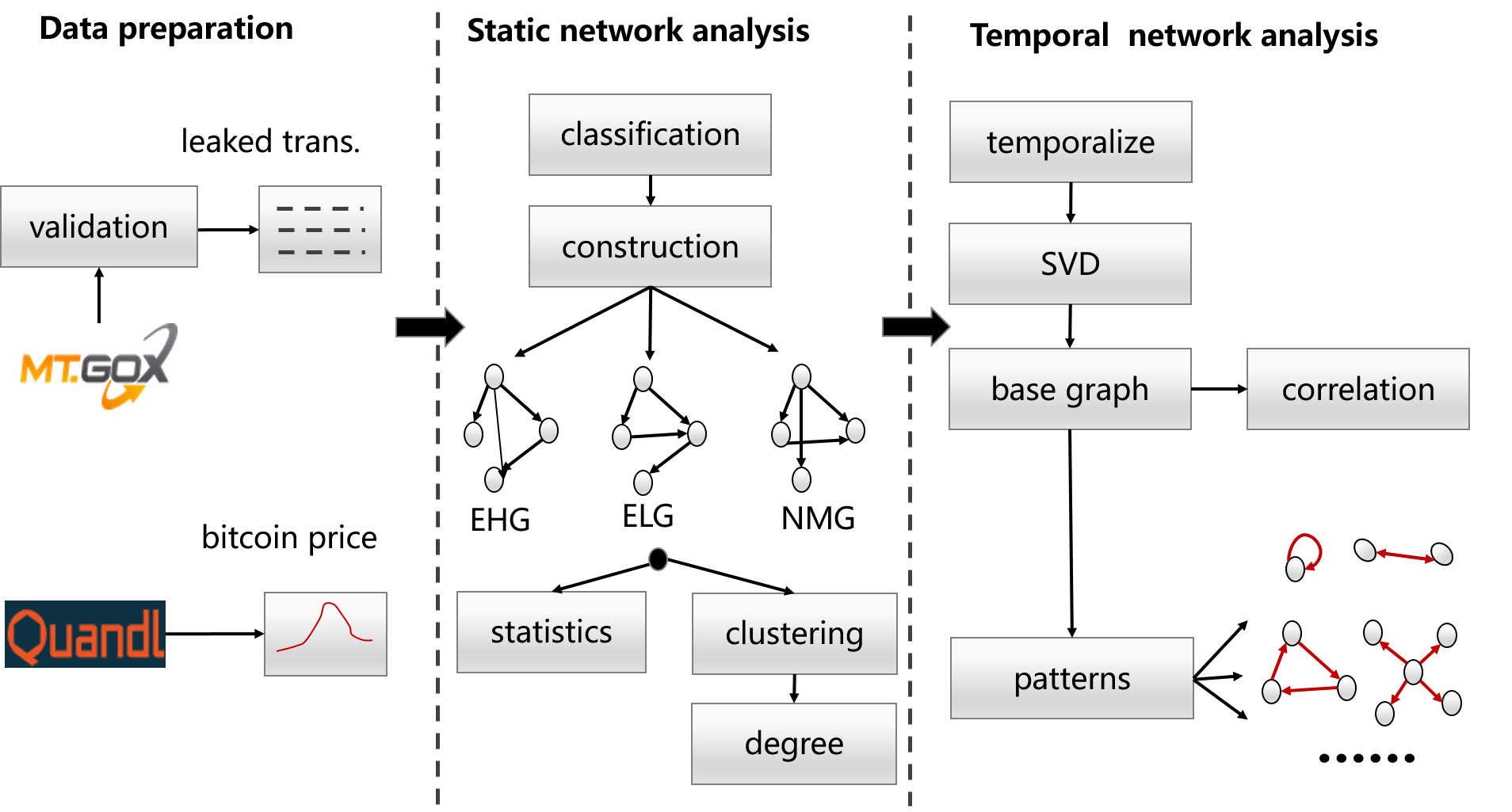}
  
  \caption{An overview of our analysis.}\label{fig_frame}
\end{figure}

Figure \ref{fig_frame} shows an overview of our analysis. We first verify the leaked data and remove many unreasonable records. Then, by comparing the transaction price with the disclosed Mt. Gox price in quandl.com, we find many abnormal transactions. By using these transactions, we divide the accounts into three categories: extreme high account (EHA), extreme low account (ELA), and normal account (NMA). Next, we construct the extreme high graph (EHG), extreme low graph (ELG) and normal graph (NMG) by seeing the accounts as nodes and transactions as edges. we conduct various graph structure analysis on EHG, ELG, and NMG, such as nodes and edges classification, measuring graph clusters and degree distribution. Such investigation leads to new observations and findings. For example, the abnormal accounts (i.e., EHA and ELA) might be controlled by the exchange and used to provide liquidity and fake volume for the exchange. Finally, by dividing the graphs into daily snapshots and reconstructing it in a matrix, we extract some base graphs through singular value decomposition (SVD). By doing this, we find that the abnormal accounts' transactions strongly related to the Bitcoin price. Furthermore, we find many strange transaction patterns (such as self-loop, bi-direction, triangle etc.) within abnormal accounts. These patterns are considered as evidence of market manipulation in the exchange.  

In summary, we make the following major contributions. 
\begin{itemize}
  \item To the best of our knowledge, it is the first study on market manipulation of cryptocurrency via graph analysis and SVD. Besides, we prove the effectiveness of the method by applying to the leaked Mt. Gox transaction data. 

  \item We obtain many new observations and findings by characterizing the activities of different accounts (i.e., static network analysis) and adopting SVD on the daily snapshots of the graphs (i.e., temporal network analysis). These findings convinced us that there are many market manipulation behaviors in the exchange. 

  \item We detect many market manipulation patterns which have never been reported in this area. These patterns are strong evidence of market manipulation and can help investors and regulators to recognize the dark side and its severity of the market. 
\end{itemize}

The rest of the paper is organized as follows. After introducing the data set in Section \ref{data}, we detail the static network analysis in Section \ref{static_analysis} and the temporal network analysis in Section \ref{sec_price_ana}. Finally, we provide some related works in Section \ref{relatedwork} and conclude the paper in Section \ref{conclusion}.


\section{Data Set}\label{data}

\begin{table*}
\caption{A segment of the leaked data.}
\centering
\begin{tabular}{ccccccccc}
\hline
Trade\_Id&    Date&    User\_Id&        Type&    Currency&    Bitcoins&    Money&    User\_Country&    User\_State\\
\hline

1380587338975940&    2013/10/1  0:28:58&    125439&        buy&    USD&    0.5&    71.69169&    US&    NC\\
1380587338975940&    2013/10/1  0:28:58&    295701&        sell&    USD&    0.5&    71.69169&    CA&    QC\\
1380739642844790&    2013/10/2  18:47:22&    609336&        buy&    USD&    0.26177217&    33.96631&    US&    PA\\
1380739642844790&    2013/10/2  18:47:22&    36865&        sell&    USD&    0.26177217&    33.96631&    US&    CA\\
\hline
\end{tabular}
\label{tab_seg}
\end{table*}



In early 2014, the transaction history from April 2011 to November 2013 of Mt. Gox was leaked in the form of CSV files. Table \ref{tab_seg} reports a segment of the leaked data recorded on 2013/10/01. Two rows with the same \emph{Trade\_Id} indicating a complete transaction from the seller (\emph{Type=sell}) to the buyer (\emph{Type=buy}). The volume of the transaction is recorded in \emph{Bitcoins} and the turnover in \emph{Money}, thus the real-time price of Bitcoin at the transaction moment is \emph{Money/Bitcoins}. Each user has a unique identity (\emph{User\_Id}) with the FIPS location codes recorded in the country (\emph{User\_Country}) and state (\emph{User\_State}) fields. There are some other attributes (e.g., transaction fees) not included in the table, as they are not used in this study. 

\textbf{Data Cleaning.} As there are many duplicate entries in the leaked data, we adopt a similar way for data cleaning as the previous studies\cite{gandal2018price,feder2018impact}. Specifically, we use the combination of the four key fields: date, user ID, type, and Bitcoins to remove duplicated entries (de-duplication strategy 2 in \cite{feder2018impact}). After this step, we remove all the single row transaction to make sure that each transaction has the corresponding buyer and seller (i.e., a completed transaction). Then, we remove all duplicated complete transactions. By doing this, the data narrows from approximately 18 million rows to 13.5 million rows (i.e, 6.7 million completed transactions). This method is more strict than the method in \cite{gandal2018price} as complete transactions with the same trade\_id  are treated as duplicates. We adopt a more strict method in the hope of providing more reliable results.

\textbf{Advantages.} The leaked Mt. Gox data has many advantages in understanding the transaction behaviors in cryptocurrency and its influence on the price. First of all, Mt. Gox was the dominant exchange and Bitcoin has been the main cryptocurrency during the period, thus analyzing the cryptocurrency market based on this data set is more reliable and representative. Second, these data are much more finely grained than data extracted from the blockchain since most trading activity is recorded only in the exchange. Furthermore, users can be identified by their accounts in the leaked data while it is hard in blockchain to identify a user because of its anonymous mechanism.    


\section{Static Network Analysis}\label{static_analysis}

\subsection{Account Classification}
Before delving deeper into the Mt. Gox leaked data, we check the Bitcoin exchange price of each transaction (i.e., Money/Bitcoin) to inspect whether it falls between the highest and lowest exchange price of the disclosed price on the same day. To this end, we first download all the Bitcoin exchange rate (BTC vs. USD) on Mt. Gox from quandl.com (we call this \emph{reference price}). Then, we compare the exchange price of each transaction with the reference price. Surprisingly, we find that there are some abnormal transactions which have a very high or low exchange price. For example, on 2013/08/30, a transaction (trade\_ID=1377875127221631) had an exchange price of \$49,338.4/BTC, and another transaction (trade\_ID=1377876535345547) had an exchange price of only \$0.81/BTC, whereas, on the same day, the highest and lowest exchange price in the download data are \$142.76/BTC and \$128.56/BTC respectively.  

These transactions are abnormal, as the exchange price is clearly out of the reasonable range. In order to distinguish the transaction behavior of different accounts and its influence on the price, we divide all the accounts into three categories: extremely high account (EHA), extremely low account (ELA) and normal account (NMA). As a first step, we apply a simple approach to identify an \emph{abnormal} transaction. For this, suppose the highest and the lowest reference price on day $t$ is $H_t$ and $L_t$, we regard an transaction with real-time price larger than $1.5\times H_t$ as an extremely high price transaction (EHT) and with real-time price lower than $0.5\times L_t$ as an extremely low price transaction (ELT). Both kinds of transactions are referred to as abnormal transactions (ABTs). Please note that we use $(0.5\times L_t,1.5\times H_t)$ instead of $(L_t,H_t)$ to identify an abnormal transaction because there are many exchanges (thus many reference price) at the same time and we cannot make sure the reference price is the real price of the exchange. However, the parameter 0.5 and 1.5 is enough to exclude any normal transaction. Finally, an account is an EHA if it has at least one extremely high price transaction and an ELA if it has at least one extremely low price transaction. Both EHAs and ELAs are referred to as abnormal accounts (ABA). Please note that abnormal accounts could be both an EHA and an ELA if it involves both EHT and ELT. NMA is an account involved in no abnormal transactions, that is to say, all involved transactions are normal transactions (NMT). 

Table \ref{tab_stat_account} shows the number of accounts and all kinds of transactions for each category of accounts. Four observations can be made from the table: 1) there are 14916 abnormal accounts, which account for 12.5\% (14916/119343) of all the accounts (please note that the number of ABA is not the sum of the number of EHA and ELA due to the existence of accounts contained in both categories); 2) the proportion of abnormal transactions (\#ABT) among ABAs accounts for 2.8\% ($\approx$194790/6775117); 3) the number of normal transactions among ABAs (3025992-194790=2831202) account for more than 41\% (2831202/6775117) of all transactions; and 4) the sum of the number of transactions (\#Tx) among ABAs and NMAs is far less than the number of all transactions, thus many transactions occurred between ABA and NMA.  

Based on these observations, one can confirm that the abnormal transactions do not occur by accident (observation 2) and the abnormal accounts behave normally in most of their times (observation 3). Thus, the existence of the abnormal accounts must have a certain special purpose. One of the most likely purposes is for providing liquidity (observation 4, Section \ref{graph_res}).  Considering the analysis on the recent cryptocurrency market of a trader and investor, which report that in some exchanges most of their disclosed trading volume are fake~\cite{fakevolume}, another possible purpose for these accounts is for fake volume. Besides, price manipulation is also a likely purpose (Section \ref{sec_price_ana}). In fact, we find that the abnormal transactions are greatly correlated with the Bitcoin exchange price and there are many abnormal patterns in the transactions.


\begin{table}
\caption{Statics of Accounts and Transactions.}
\centering
\begin{tabular}{cccccc}
\hline

Category & \#accounts  & \#Tx & \#ABT & \#EHT & \#ELT \\ 
\hline
  EHA &   10702 & 1406850 &  179701 &  138743 &   40958 \\ 
  ELA &    5835 & 2486807 &   85784 &   29737 &   56047 \\ 
  ABA &   14916 & 3025992 &  194790 &  138743 &   56047 \\ 
  NMA &  104427 & 812865 &      0 &      0 &      0  \\ 
  \hline
 All &  119343 & 6775117 &  194790 &  138743 &   56047  \\ 
   \hline
\end{tabular}
\label{tab_stat_account}
\end{table}


\subsection{Graph Construction}\label{graph_construct}
As each transaction contains a buyer and a seller, we can easily construct a directed graph from the records by considering each account as a node. Specifically, we present the definition of the constructed graph $G$ as follows.

\textbf{Graph Definition.} $G=(V,E,w)$, where $V$ is a set of nodes represent users (denoted by user ID) in the leaked data, $E$ is a set of edges with each represents an \emph{ordered} pair of nodes and $w$ is the function associating each edge to a weight. Each pair indicates that there was at least one transaction between users $u$ (seller) and $v$ (buyer) in the whole dataset. $w:E\rightarrow \mathbb{R}_+$ maps each edge with a weight, which is the total amount of Bitcoins transferred along the edge by one or more transactions.   

In the remainder of this paper, we use the term \emph{account}, \emph{user} and \emph{node} interchangeably. To better compare network characteristics, we construct three graphs according to the nodes' categories as follows:
\begin{itemize}
\item EHG. The graph that all nodes are EHAs. 
\item ELG. The graph that all nodes are ELAs. 
\item NMG. The graph that all nodes are NMAs. 
\end{itemize}

To construct the graph we adopt the following steps. Since each complete transaction has both a buy and sell record (has the same transaction ID) after data validation, we first construct a set of tuples $(S,B,v,t,l)$ from every complete transaction, where $S$ and $B$ represents the seller and buyer (denoted by user ID), $v$ is the corresponding amount of the transaction in Bitcoin, $t$ is the transaction time and $l$ is a label indicating the category of the transaction (i.e., EHT, ELT or NMT). We call this set as \emph{transaction tuple}, as each tuple corresponds to a unique transaction. Based on the transaction tuple, the aforementioned graphs are easy to construct. For example, to construct the EHG, we select all the tuples in which both the seller and the buyer are EHAs and sum the $v$ entry grouped by $S$ and $B$. Then, the generated new tuples $(S,B,v)$ is the EHG. Other graphs are constructed as the same except by selecting different tuples according to the nodes' category.


{}

\subsection{Graph Analysis} \label{graph_res}
\begin{figure}[htbp]
\centering

  \subfloat[EHG]{%
    \includegraphics[width=.16\textwidth]{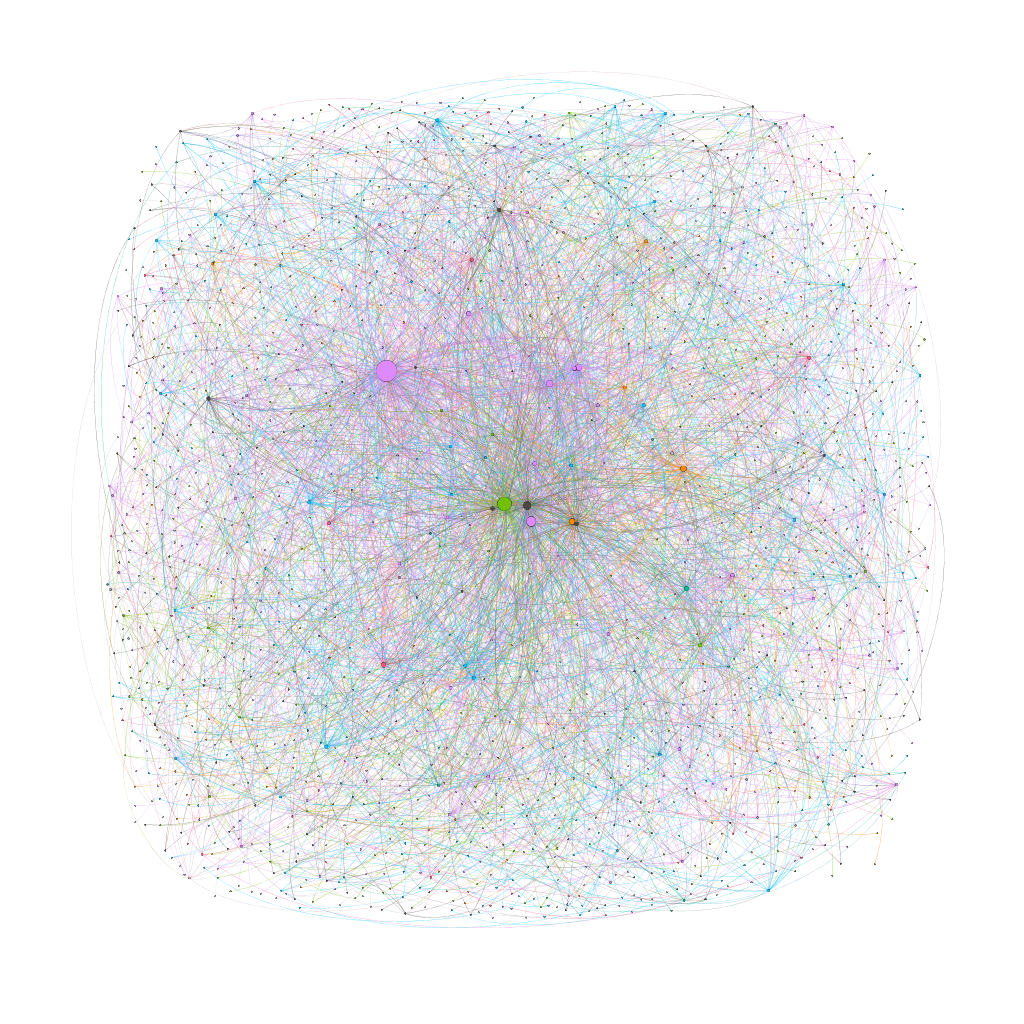}}\hfill
     \vspace*{1mm}
  \subfloat[ELG]{%
    \includegraphics[width=.16\textwidth]{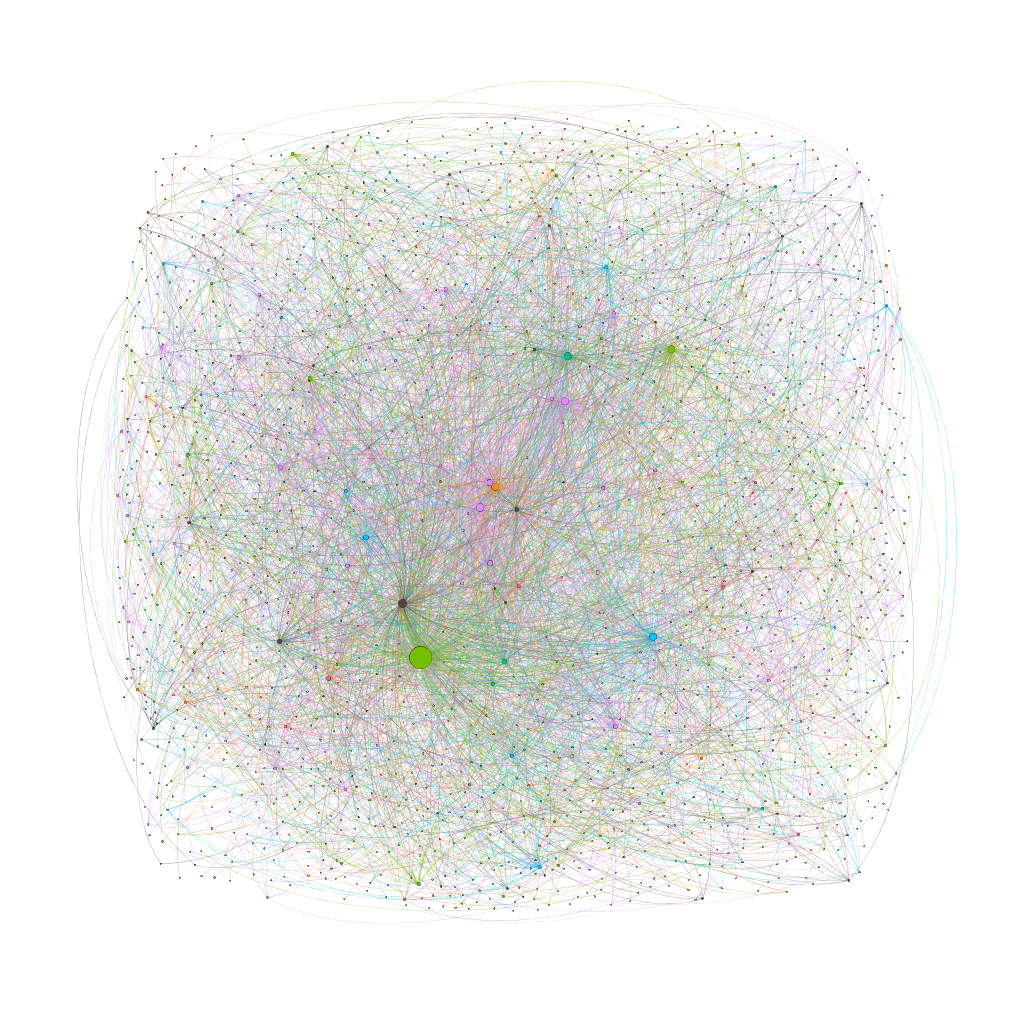}}\hfill
  \subfloat[NMG]{%
    \includegraphics[width=.16\textwidth]{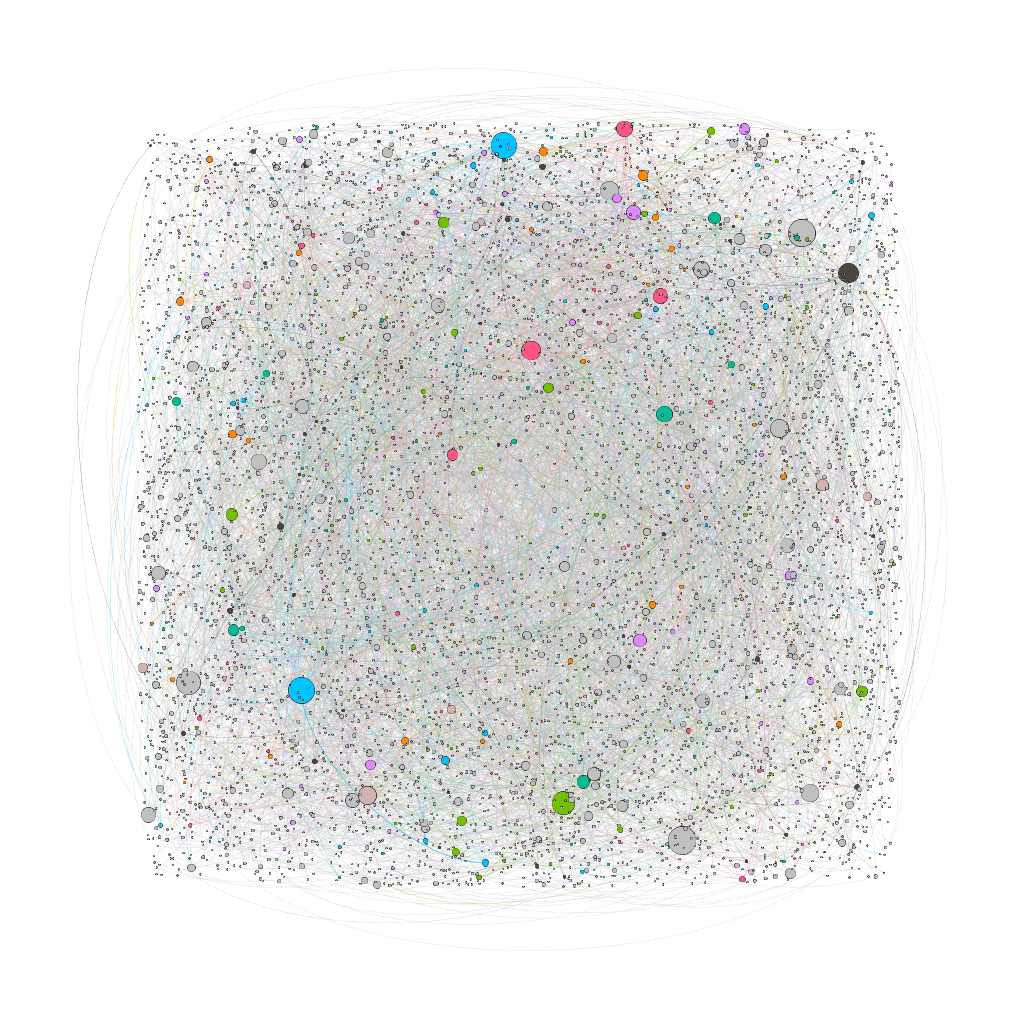}}\

  \caption{Visualization of EHG, ELG, and NMG. For the ease of illustration, we randomly select 5000 edges from each graph to draw the figure.}\label{fig_all}

\end{figure}

This subsection investigates the constructed graphs from various metrics in graph analysis. Figure \ref{fig_all} shows the three graphs. We can find that there are more nodes in NMG, indicating the NMG is more sparse in connection (note that we select 5,000 edges for each graph). We investigate the statistics and metrics in the following. 

\vspace{0.5cm}
Table \ref{tab_stat} shows all the statistics and metrics for each constructed graph. For comparison, we also constructed the abnormal graph (i.e., the graph of all abnormal accounts, ABG) and the complete graph (i.e, the graph of all accounts, CG). In the following, we first introduce the statistics or metrics and then detail the observations. 

The number of nodes in each graph is the number of accounts in each category, which is in accordance with the statistics in Table \ref{tab_stat_account}. The only exception is that the number of nodes in NMG is less than the number of NMA, because some normal accounts interact with abnormal accounts, thus it is not included the NMG.     
\begin{table}[htbp]
\caption{Statics of graphs.}
\centering
\begin{tabular}{cccccccccc}
\hline

graph & \# nodes   & \# edges & cluster & avg. degree & avg. wgt. degree\\ 
\hline
  EHG &   10702  &  212900 & 0.30& 19.89 & 505.43  \\ 
  ELG &    5835  &  413881& 0.42 & 70.93 & 3107.68   \\ 
  ABG &   14916  &  612885 & 0.31& 41.09 & 1439.04 \\ 
  NMG &  86457   & 655882& 0.03& 7.59& 76.21  \\ 
  CG &  119343   & 2682719  & 0.28& 22.48 & 426.54 \\ 
   \hline

\end{tabular}
\label{tab_stat}
\end{table}

An edge in the graph indicates a ``channel'' between two accounts for buying or selling Bitcoin. As can be seen from the table, the number of edges in each graph is far less than the number of transactions, which means that many channels are used more than one times. Another notable result is that the summation of the number of edges in ABG and NMG is greatly less than the number of edges in the CG. This result indicates that many edges are the channels between normal and abnormal accounts and is evidence that the abnormal accounts provide liquidity in the exchange. The number of edges in ABG is slightly larger than the sum of the number of edges in EHG and ELG since there are some edges connecting EHAs and ELAs.  

We compute the clustering coefficient of all the graphs in column 4 of Table~\ref{tab_stat}. As can be seen, the clustering coefficients are extremely different among EHG, ELG, and NMG. The large clustering coefficients (i.e., 0.3 in EHG and 0.42 in ELG) revealing that if two abnormal accounts $A, B$ trade with abnormal account $C$, $A$ and $B$ are very likely to trade with each other. In other words, the abnormal accounts are likely to form triangles through transactions. Conversely, the clustering coefficient of NMG is very small (i.e., 0.03), which indicates a normal situation as the probability of three normal accounts forming a triangle is very small. This result indicates that the abnormal accounts behave strangely and herald the existence of market manipulation in the exchange. 

\begin{figure}[htbp]
\centering
  \subfloat[EHG]{%
    \includegraphics[width=.16\textwidth]{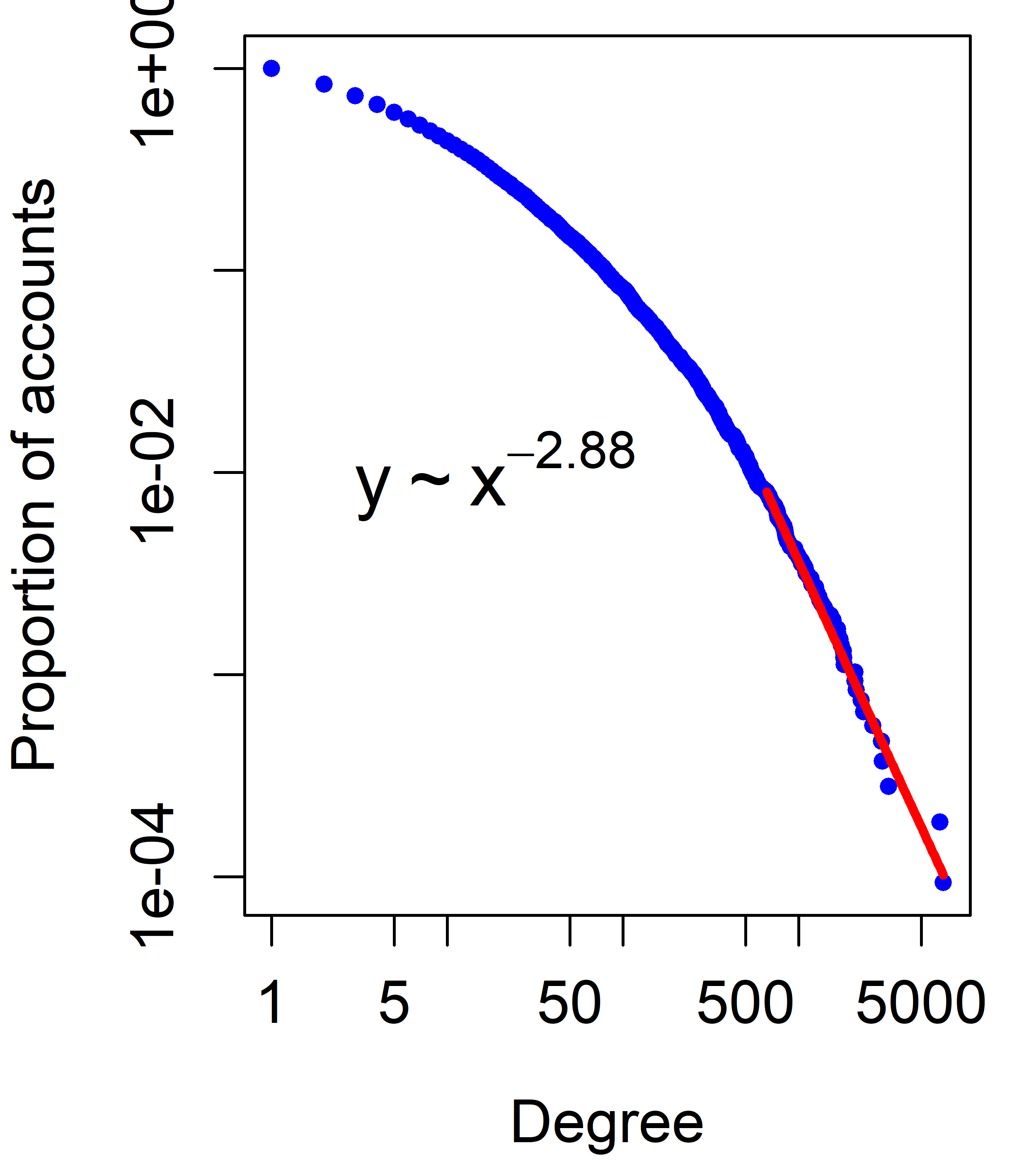}}\hfill
     \vspace*{1mm}
  \subfloat[ELG]{%
    \includegraphics[width=.16\textwidth]{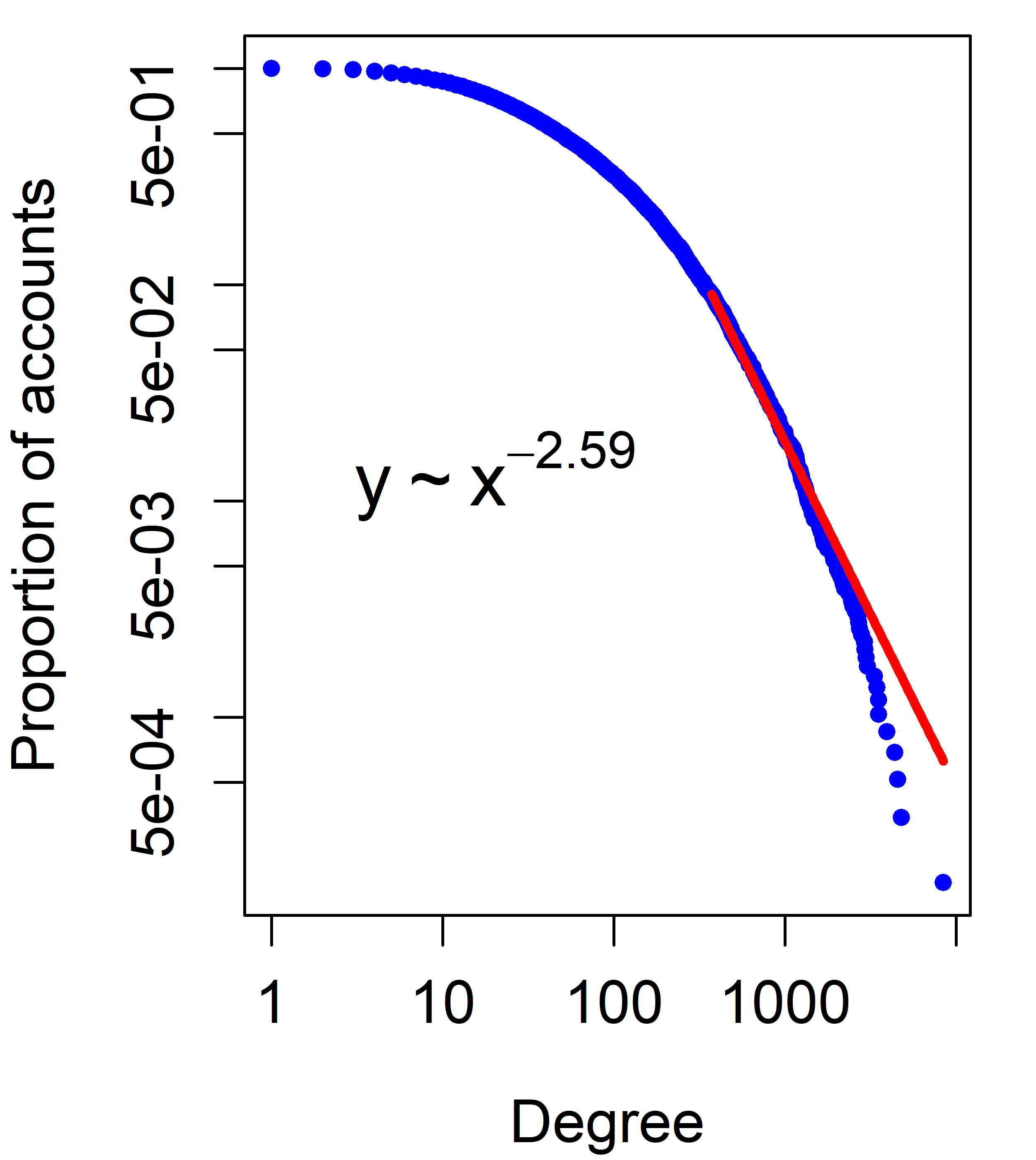}}\hfill
  \subfloat[NMG]{%
    \includegraphics[width=.16\textwidth]{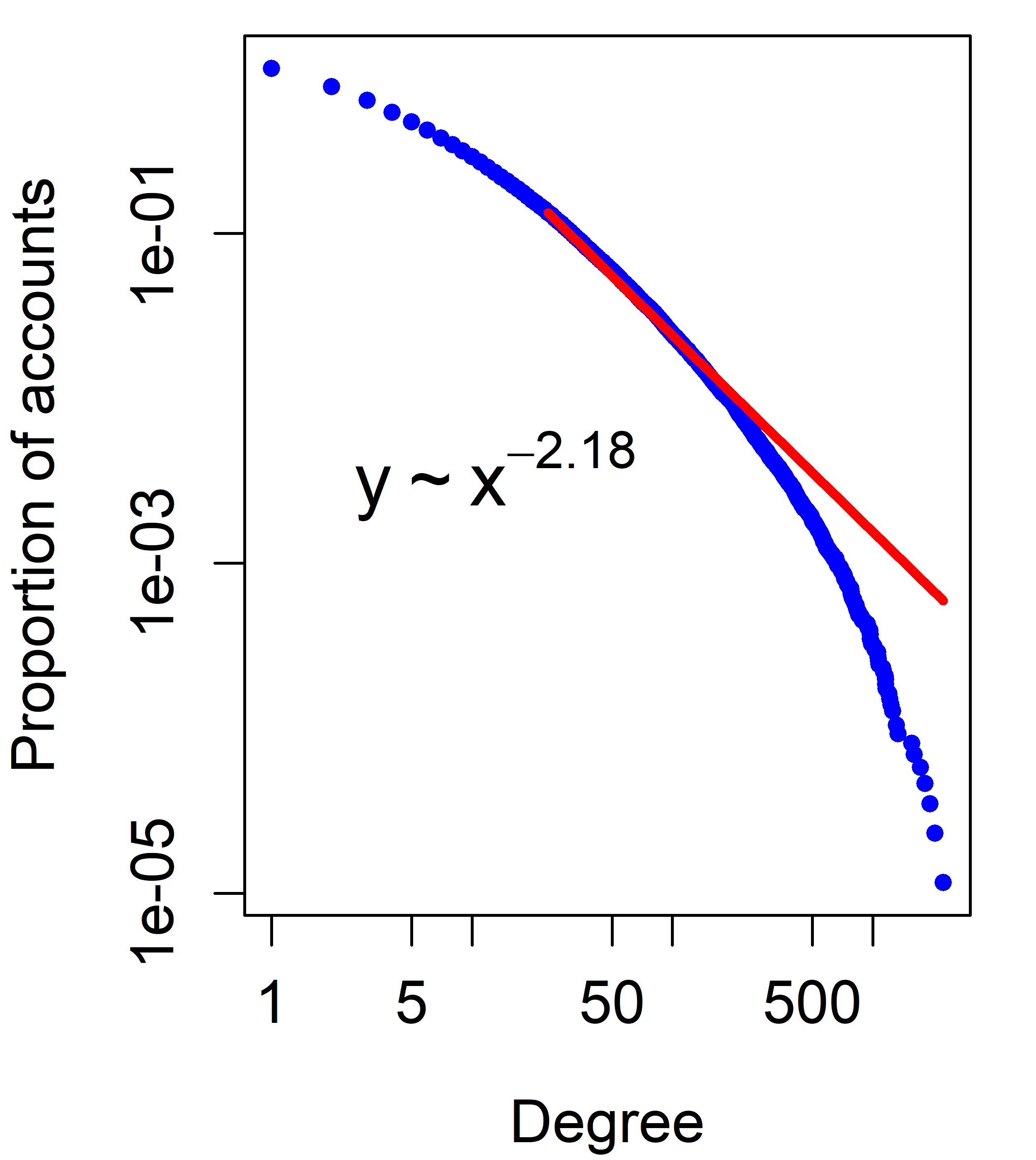}}

  \caption{Degree distribution of EHG, ELG and NMG.}\label{fig_degree_dis}
\end{figure}

The degree of a node is the number of edges connecting to the node. In our case, the degree of a node indicates the number of accounts trading with that node. Figure~\ref{fig_degree_dis} shows the degree distribution of all the three graphs, all of which approximately follows the power law distribution, meaning that there are few large-degree nodes and many small-degree nodes. We estimate the parameters by using the free statistical software R\cite{R} and the contributed package~\cite{poweRlaw} and plot the fitting line $y\sim x^{-\alpha}$ for each distribution in red. The smaller the $\alpha$, the more variable of nodes' degree. Thus, the abnormal accounts show less variable as compared with normal accounts. The result may be due to the abnormal accounts are controlled by the same organizations.         

Column 5 and 6 in Table \ref{tab_stat} show the average degree and the weighted average degree of the graphs. The large average degrees of EHG and ELG indicate that the abnormal accounts are used more frequently than normal accounts. The weighted degree is computed by setting the transaction volume (i.e., Bitcoin) as the weight, thus the average weighted degree represents the average transaction volume for each edge. As can be seen, the average weighted degree of ELG is far larger than it of EHG, one possible reason that the exchange price of transactions in ELG is relatively low, thus the transaction volume is large. Whatever the reason is, an obvious fact remains that the average weighted degree of EHG and ELG are larger than that of NMG, which means the edges between abnormal accounts transfer more Bitcoin than edges between normal accounts.

Based on the results and analysis discussed above, we summarize the findings as follows:

\begin{itemize}
\item  \textbf{Finding 1.} There are some abnormal accounts (12.5\%) which trading with very high or low exchange price in some transactions. We consider these accounts abnormal and under control by the exchange for two reasons: 1) the proportion of the abnormal transactions account for 2.8\%, thus it is not occurred by accident; 2) the abnormal exchange price is impossible to appear on ordinary users.

\item \textbf{Finding 2.} Many seemingly normal transactions occurred between abnormal accounts ( $>$ 41\%). There are two possible purposes for these transactions: 1) these transactions are the fake volume that used to create an illusion of active trading; 2) to provide liquidity for the exchange.  

\item \textbf{Finding 3.} The graphs of abnormal accounts have very large clustering coefficients. One possible reason is that these accounts are controlled by one organization, and thus the trade is not completely random. 

\end{itemize}

These findings indicate that the exchange was likely involved in trading manipulation. As the exchange price is the key factor of trading, in the following section, we will discuss the possibility of price manipulation of the exchange.







\section{Temporal Network Analysis}\label{sec_price_ana}
As discussed above, the transaction network of abnormal accounts (i.e., EHG and ELG) show a great difference from the NMG. We want to know whether these transactions have a correlation with the Bitcoin price and what kind of users and transactions (i.e., graph structure) influence the Bitcoin price greatly. To this end, we calculate the daily snapshots of the graphs by adopting the method similar to \ref{graph_construct}. To detect important changes in the graph structure, we compare successive snapshots of the graphs using singular value decomposition (SVD). The goal is to detect a set of base networks and represents each day's snapshot as a linear combination of these base networks. Unlike in Section \ref{static_analysis}, we focused our study on transaction data after 2012/12/01 in this section. There are many reasons supporting our choice. Firstly, the recent paper which proves the price manipulation of Mt. Gox uses the same transaction history \cite{gandal2018price}. Secondly, the Bitcoin price experienced a skyrocketing during this period. Thirdly, Mt. Gox was the main Bitcoin exchange during this period. Finally, more abnormal users and transactions (more than 60\%) are found after that day.  




\subsection{Extract Base Networks}
To evaluate which networks influence the price greatly, we need to construct the daily snapshots of the three graphs: EHG$_t$, ELG$_t$ and NMG$_t$. We adopt the same process to construct the graph series. First of all, we construct the \emph{aggregate} networks (i.e., EHG) based on tuples after 2012/12/01. Assume there are $n$ nodes and $L$ edges in the aggregate network, then it can be represented by a $n\times n$ weighted adjacency matrix $G$, in which there are $L$ non-zero elements. We rearrange $G$ into an $L$ long vector $g$ containing all the non-zero elements. We call this vector as \emph{edge-weight} vector. The vector describes the \emph{graph structure} of the aggregate network as each element represents a possible edge and its weight. To construct the daily snapshots of EHG$_t$ on day $t$, we recalculate the edge-weight vector $g_t$ (i.e., the graph structure on day $t$) based on transaction tuples on day $t$. Please note that we do not change the order of the vector, thus the $i$-th element of all the edge-weight vectors indicate the same edge, and it may be zero if the edge does not exist on a specific day. For $T$ snapshots, we now build the $T\times L$ graph time series matrix $X$ such that the $t$-th row of $X$ equals $g_t$. By doing this, we build a special matrix with $T$ samples and each sample represents a daily graph structure.   

To account for the variation of the daily graph structure, we normalize $X$ such that the sum of each row equals 1, and then subtract the column averages from each column. As a result, both the row and column sums in the matrix will be zero. We compute the singular value decomposition of the matrix $X$: 
\begin{equation}
X=U\Sigma V^T,
\end{equation}
where $U$ is a $T\times T$ unitary matrix, $\Sigma$ is a $T\times L$ diagonal matrix with non-negative values on the diagonal, and $V$ is a $L\times L$ unitary matrix. The non-negative values on the diagonal are \emph{sigular} values and is usually sorted in descending order. The left-singular vectors containing in the column of $U$ are a set of orthonormal eigenvectors of $XX^T$, 
and the right-singular vectors containing in the column of $V$ are a set of orthonormal eigenvectors of $X^TX$.
Since in this case $T<L$, there are only $T$ nonzero sigular values. We denote the sorted sigular values as $(\sigma_1,\cdots,\sigma_T)$, the left-sigular vectors $(\vec{u_1},\cdots,\vec{u_T})$ and the right-sigular vectors $(\vec{v_1},\cdots,\vec{v_T})$, where $\vec{u_i}$ and $\vec{v_i}$ are column vectors and subject to the following equations:
\begin{equation}
\vec{u_i}^T*\vec{u_j}=\vec{v_i}^T*\vec{v_j}=\delta_{ij}.
\end{equation}

Based on the special meaning of matrix $X$, we can interpret the singular vectors and the singular values as 1) the right-singular vectors can be seen as \emph{base networks}, and the element $v_i(l)$ (i.e., the $l$-th element of the $i$-th right-singular vector) gives the weight of the $l$-th edge in the $i$-th base network; 2) the left-singular vectors account for the temporal variation of the base networks, the $t$-th value of $\vec{u_i}$ (denotes as $u_i(t)$) provides the contribution of the $i$-th base network on day $t$; 3) the singular value $\sigma_i$, which are the square roots of the non-zero eigenvalues of both $X^TX$ and $XX^T$, indicates the overall importance of the $i$-th base network in approximating the whole matrix. Please note that the singular values are sorted in decreasing order, thus give decreasing contribution to the result. 

\subsection{Detecting Graph Structural Changes}
As the (normalized) weight of the $l$-th edge in the daily graph structure on day $t$ can be written as:
\begin{equation}
x_{tl} = \sum_{i=1}^T \sigma_i u_i(t)v_i(l),
\end{equation}
to detect graph structural changes, we need to consider two terms: $\sigma_{i}$ (i.e., the importance of the $i$-th base network) and $u_i(t)$ (i.e., the contribution of the $i$-th base network on day $t$). 

As a first glance, we consider the daily influence of the first and also the most important base network (i.e., $u_1(t)$). We want to know the correlation between the variation of $u_1(t)$ and the fluctuation of the Bitcoin exchange price. As the range of the price is $(12, 1207)$, we adopt a simple mathematical transform to make sure most of the transformed price falls in the interval $(0, 1)$. Specifically, we adopt the log transform $B(t)=log_{1000}P(t)$, where $P(t)$ is the close exchange price of Bitcoin on day $t$. Table \ref{tab_u1r} (left part) shows three commonly used correlation coefficients (i.e., Pearson, Spearman, and Kendall correlation coefficient) between $u_1(t)$ and the log-transformed price $B(t)$. The results show that the daily variation of the first base network in EHG and ELG have a very strong correlation with the Bitcoin exchange price. However, in NMG, there is no correlation between the two variables. The result indicates that the transactions made between abnormal accounts have a great influence on the Bitcoin exchange price.      

\begin{table}
\caption{ Correlation coeffcients between the left-singular vectors of the network time series matrix and the Bitcoin exchange price.}
\centering
\begin{tabular}{c|ccc||ccc}
\hline

\multirow{2}{*}{Graph}& \multicolumn{3}{c||}{The 1st base network} & \multicolumn{3}{c}{The Fitted 10 base networks}\\
\cline{2-7}
 &  $\rho_{\rm{P}}$&  $\rho_{\rm{S}}$ & $\rho_{\rm{K}}$ & $\rho_{\rm{P}}$&  $\rho_{\rm{S}}$& $\rho_{\rm{K}}$ \\
\hline
EHG& 0.56& 0.60&  0.44& \textbf{0.811} & \textbf{0.807}& 0.620\\

ELG& 0.58& \textbf{0.82}&  0.64 & \textbf{0.871}& \textbf{0.834} &0.652\\

NMG& 0.05& 0.15&  0.12 &0.239& 0.398 &0.289\\
\hline
\end{tabular}
\label{tab_u1r}
\end{table}

\begin{figure}[htbp]
\centering
 \includegraphics[width=0.45\textwidth]{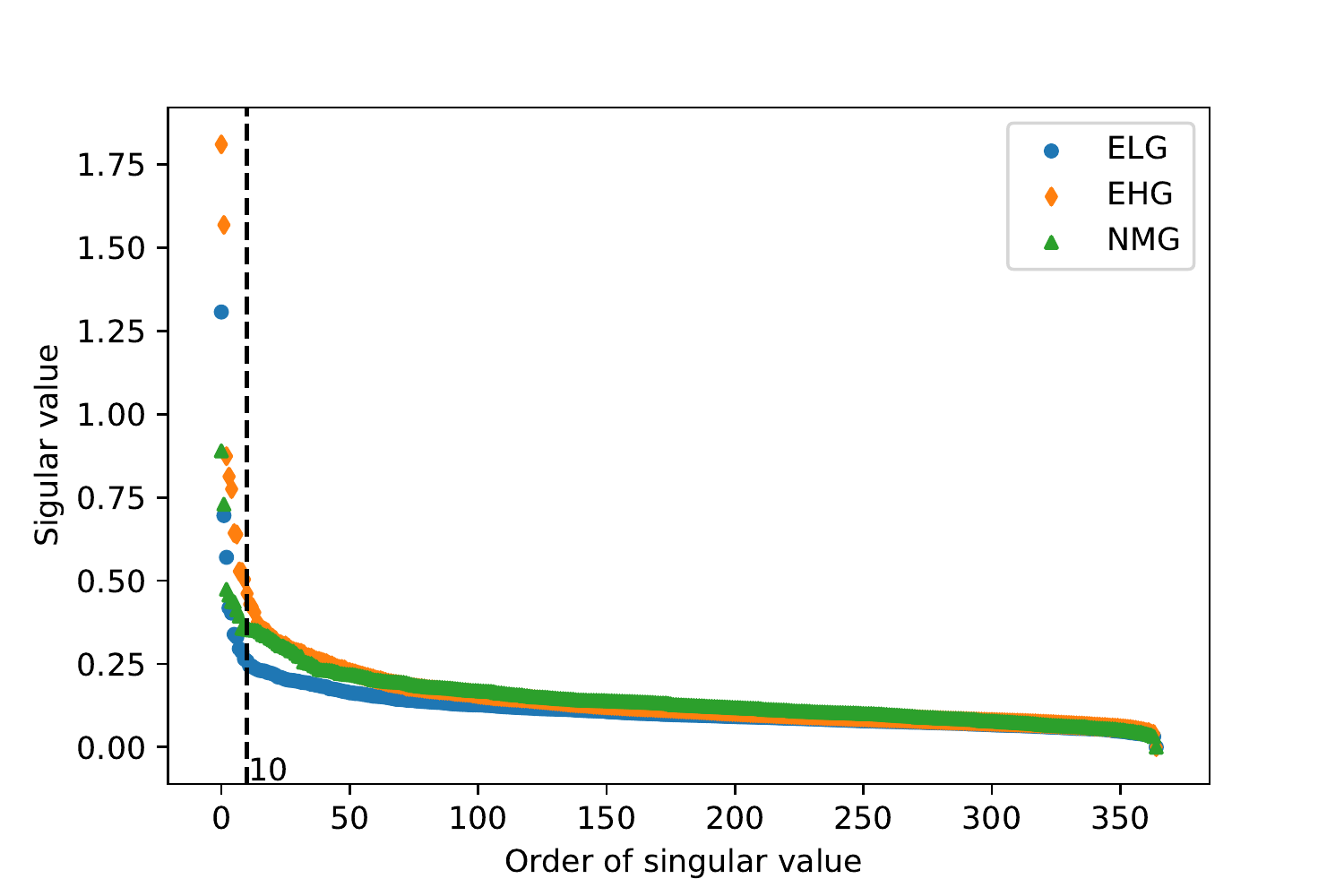}

  \caption{Sigular values in the order of its importance.}\label{sigularvalues}

\end{figure}

\vspace{0.5cm}

\begin{figure*}[thbp]
\centering

  \subfloat[EHG]{%
    \includegraphics[width=.32\textwidth]{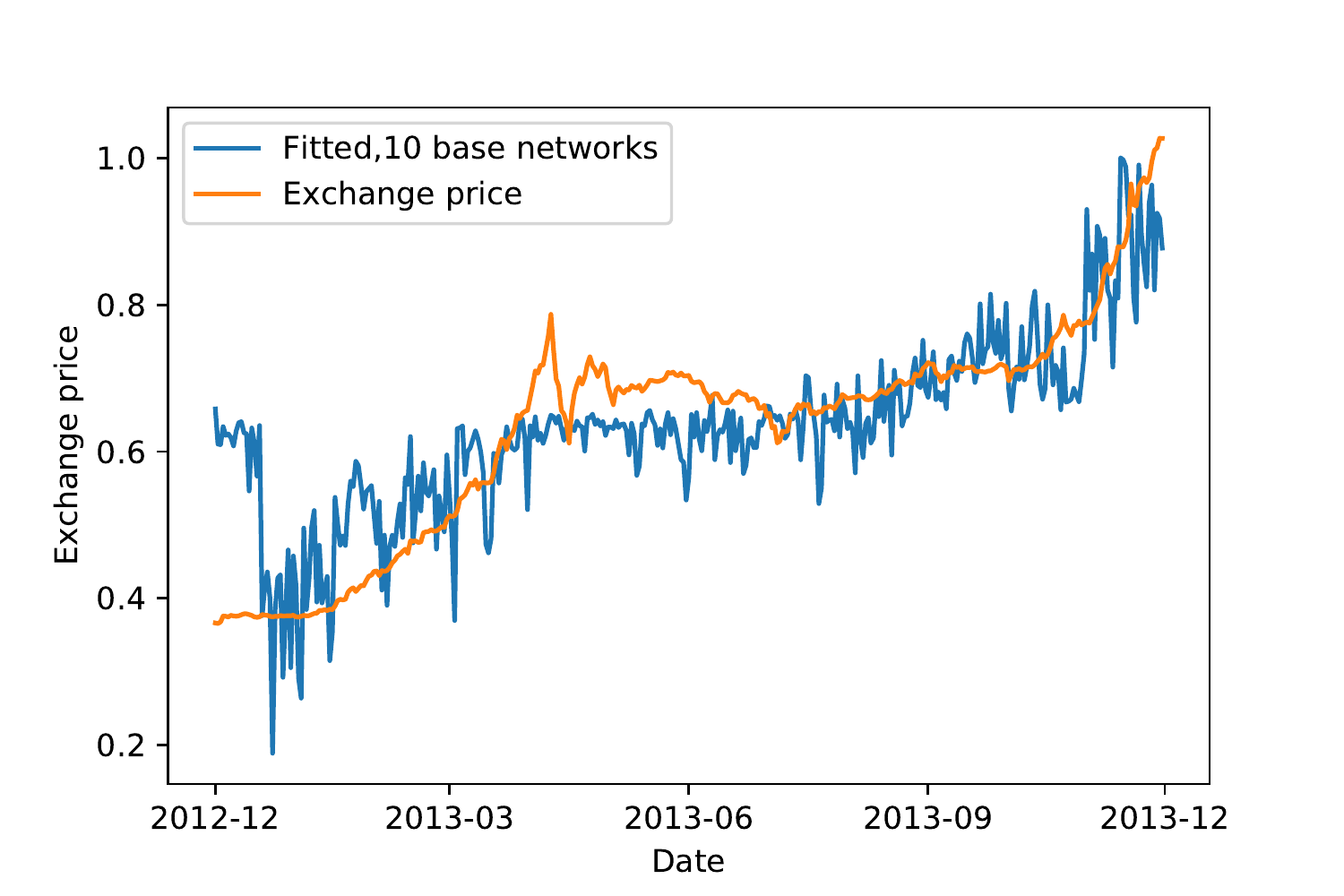}}\hfill
     \vspace*{1mm}
  \subfloat[ELG]{%
    \includegraphics[width=.32\textwidth]{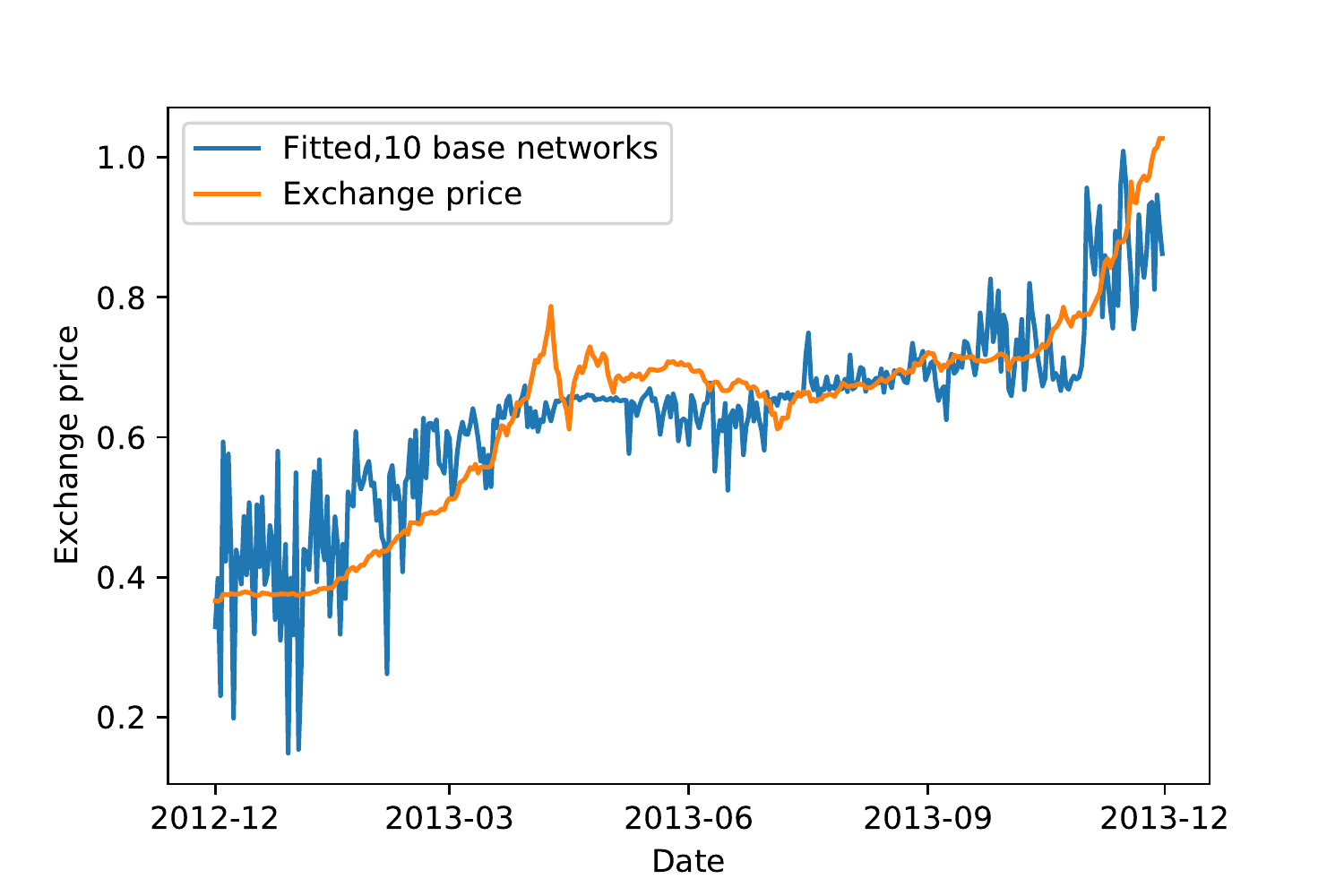}}\hfill
  \subfloat[NMG]{%
    \includegraphics[width=.32\textwidth]{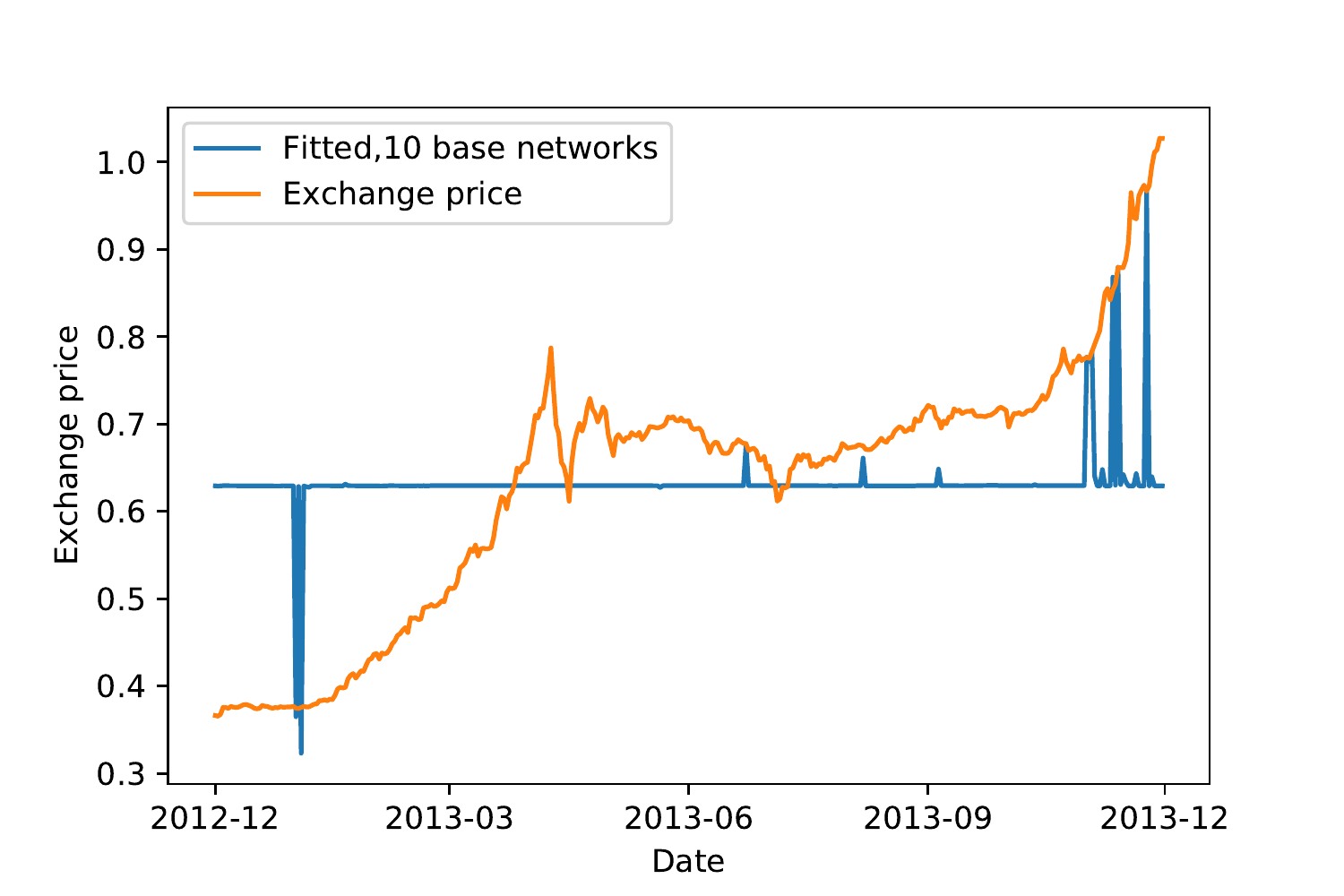}}\

  \caption{Approximate the log-transformed Bitcoin price with the linear combination of the selected base networks of EHG, ELG, and NMG.}\label{fit_price}

\end{figure*}

\begin{figure*}[thbp]
\centering

  \subfloat[EHG]{%
    \includegraphics[width=.31\textwidth]{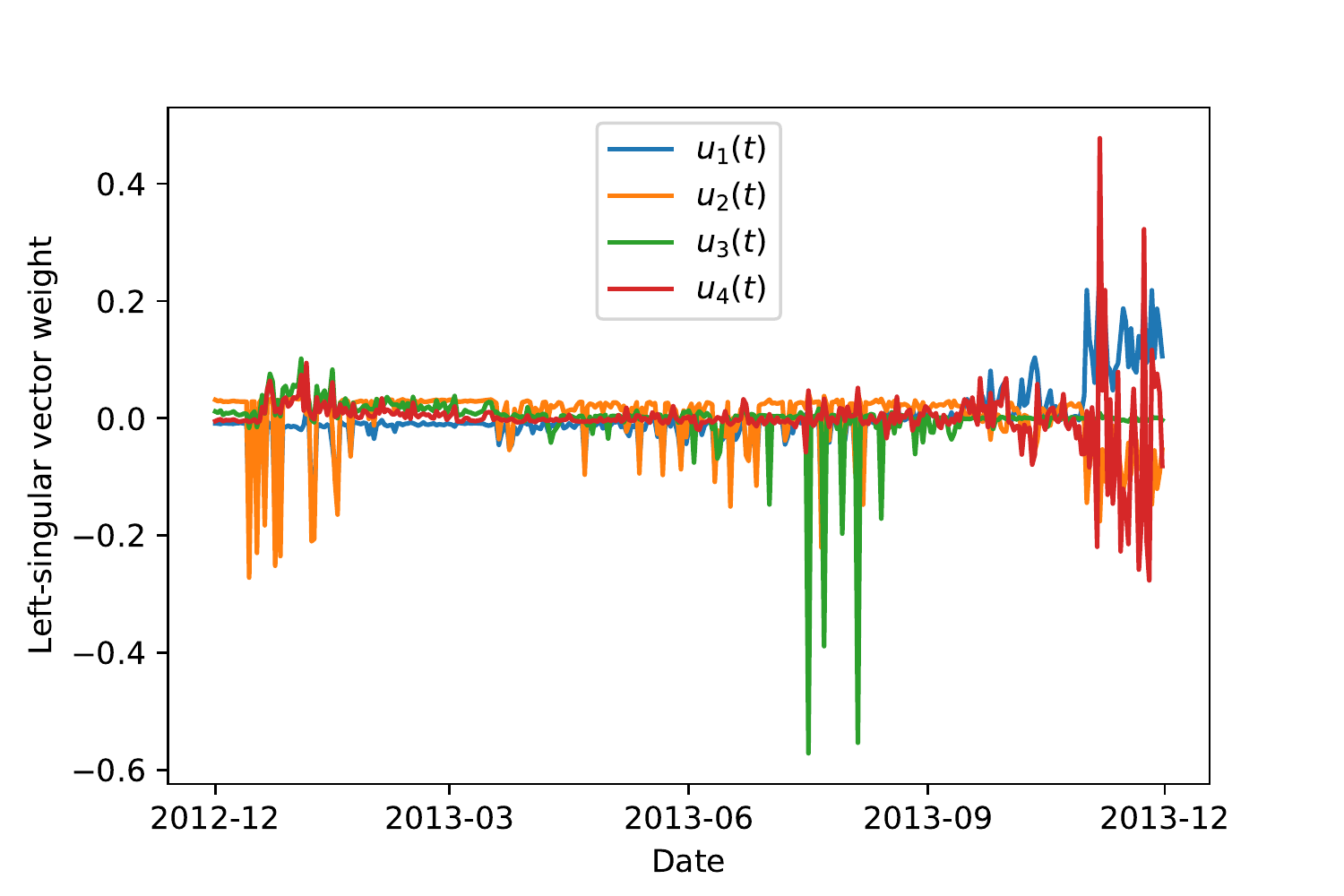}}\hfill
     \vspace*{1mm}
  \subfloat[ELG]{%
    \includegraphics[width=.31\textwidth]{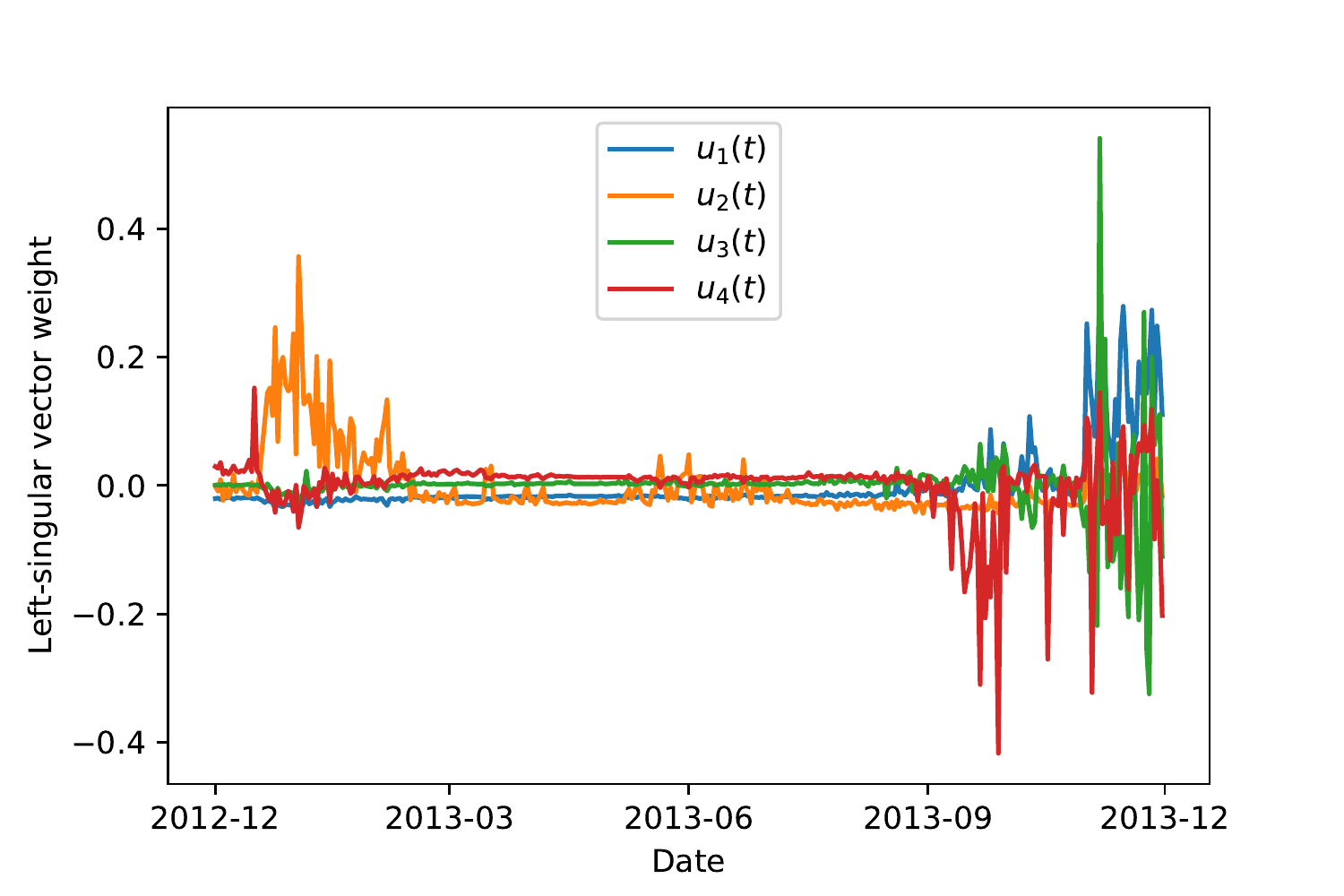}}\hfill
  \subfloat[NMG]{%
    \includegraphics[width=.31\textwidth]{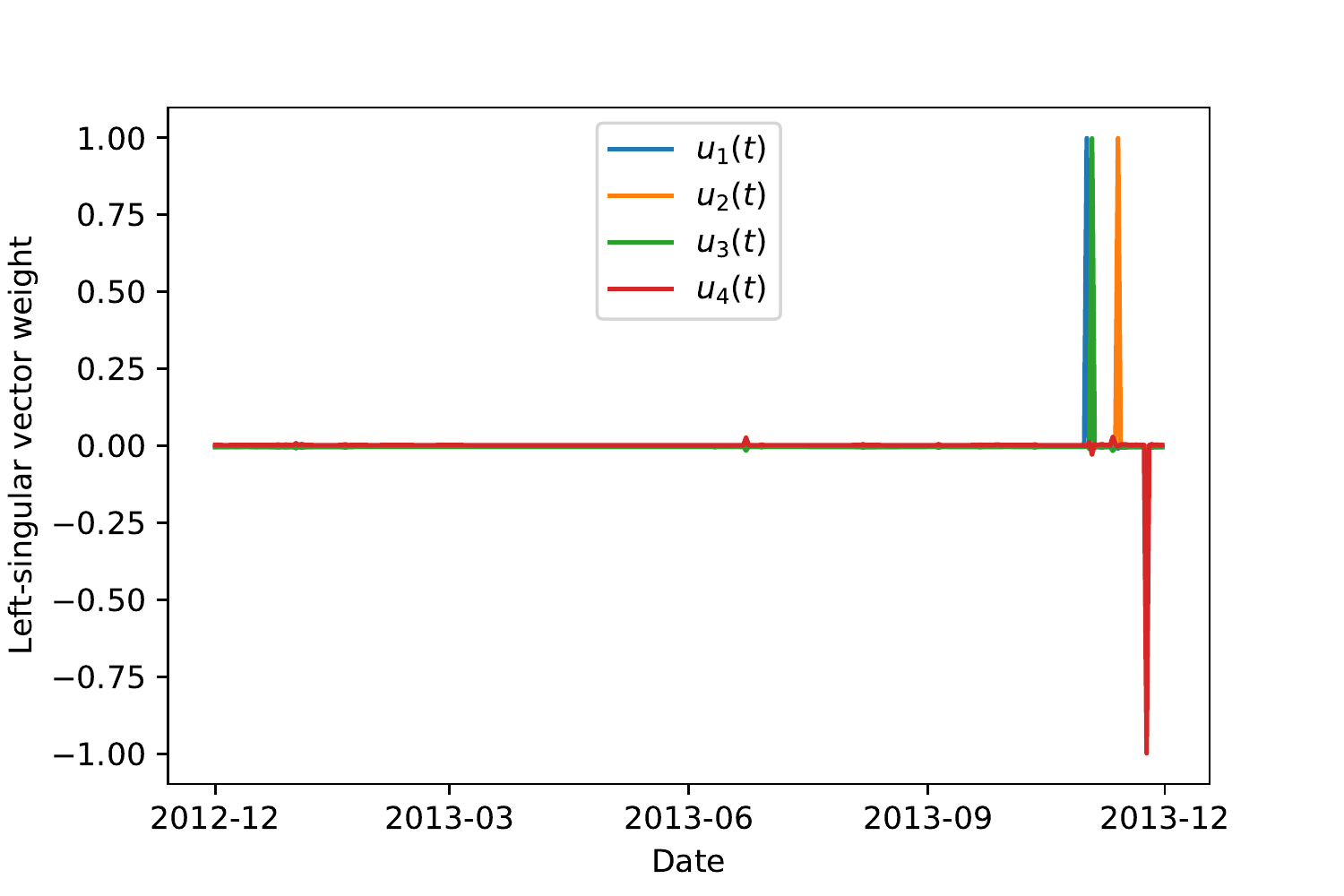}}\

  \caption{The time-varying contribution $u_i(t)$ of the first four base networks.}\label{uit}

\end{figure*}  
Motivated by this result, we want to know, to what extent, the log transfered price can be estimated with the combination of the left-sigular vectors, i.e.,
\begin{equation}\label{discompose}
B(t) \sim c_0 + \sum_{i=1}^N c_iu_i(t),
\end{equation} 
where $c_0$ is the mean of $B(t)$ and $c_i$ can be computed as the dot product of $B(t)$ and $u_i(t)$. 
As the left-singular vectors are orthonormal and span the T-dimensional linear space, $B(t)$ can be reconstructed by $u_i(t)$ when $N=T$. However, this is not what we desire in this case. The purpose of this study is to identify some important base networks and accounts that have a great influence on the Bitcoin price. To proceed, we first try to select some important base networks in the detected base networks. We draw the scree plot of the singular values as shown in Fig. \ref{sigularvalues}. As can be seen from the graph, the curve of the singular values is clearly leveling off at the right side of the dotted line (i.e., the 10th singular value). Thus, we select the first 10 base networks for the following analysis.

Before analyzing accounts in the selected base networks, we approximate $B(t)$ with the selected networks. To evaluate the fitting effect, we calculate the correlation coefficients between the fitted price series and $B(t)$. The right part of Table \ref{tab_u1r} shows the correlation coefficients. Surprisingly, the three correlation coefficients are greatly enhanced as compared with the first left-singular vector. Especially, the Pearson correlation coefficient between ELG and $B(t)$ is 0.87, while only 0.24 between NMG and $B(t)$. The great difference indicates a strong correlation between abnormal accounts' transactions and the Bitcoin exchange price, which is a strong evidence of the price manipulation in Mt. Gox.

Figure \ref{fit_price} shows the trends of $B(t)$ and the fitted price. As can be seen from the graph, though the shape of the peak in April of 2013 is missed, the trends of $B(t)$ has been grasped by the selected base networks of EHG and ELG, whereas the base networks in the NMG have no effect in grasping the trend.

To show the structure variation of the networks, we draw the time-varying contribution $u_i(t)$ of the first four base networks in Fig. \ref{uit}. 
In most cases, $u_i(t)$ exhibit a few abrupt changes, partitioning the history of the transaction into separate time periods. The most notable abrupt changes are in December of 2012 when the Bitcoin exchange price is very smooth and the November of 2013 when the price skyrocketing. During the two periods, the effects of the first four base networks of EHG and ELG are both significant, however, the base networks in NMG have no distinct effect during the smooth period and show effect only a few days during the skyrocketing period.

\begin{figure*}[htbp]
\centering

  \subfloat[Self-Loop]{%
  \label{Self-Loop}
    \includegraphics[width=.26\textwidth,height=.26\textwidth]{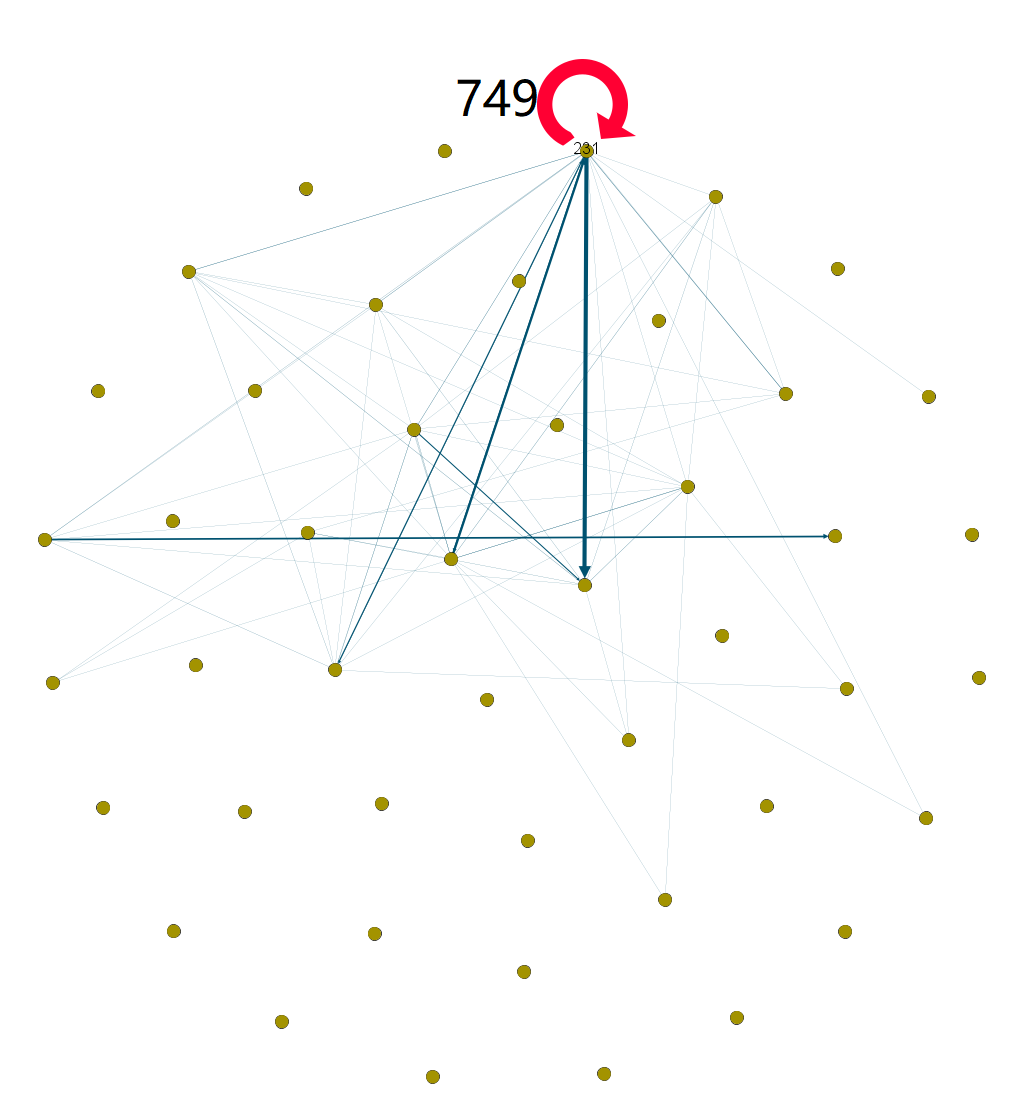}}\hfill
     \vspace*{1mm}
  \subfloat[Unidirection]{%
  \label{Unidirection}
    \includegraphics[width=.26\textwidth,height=.26\textwidth]{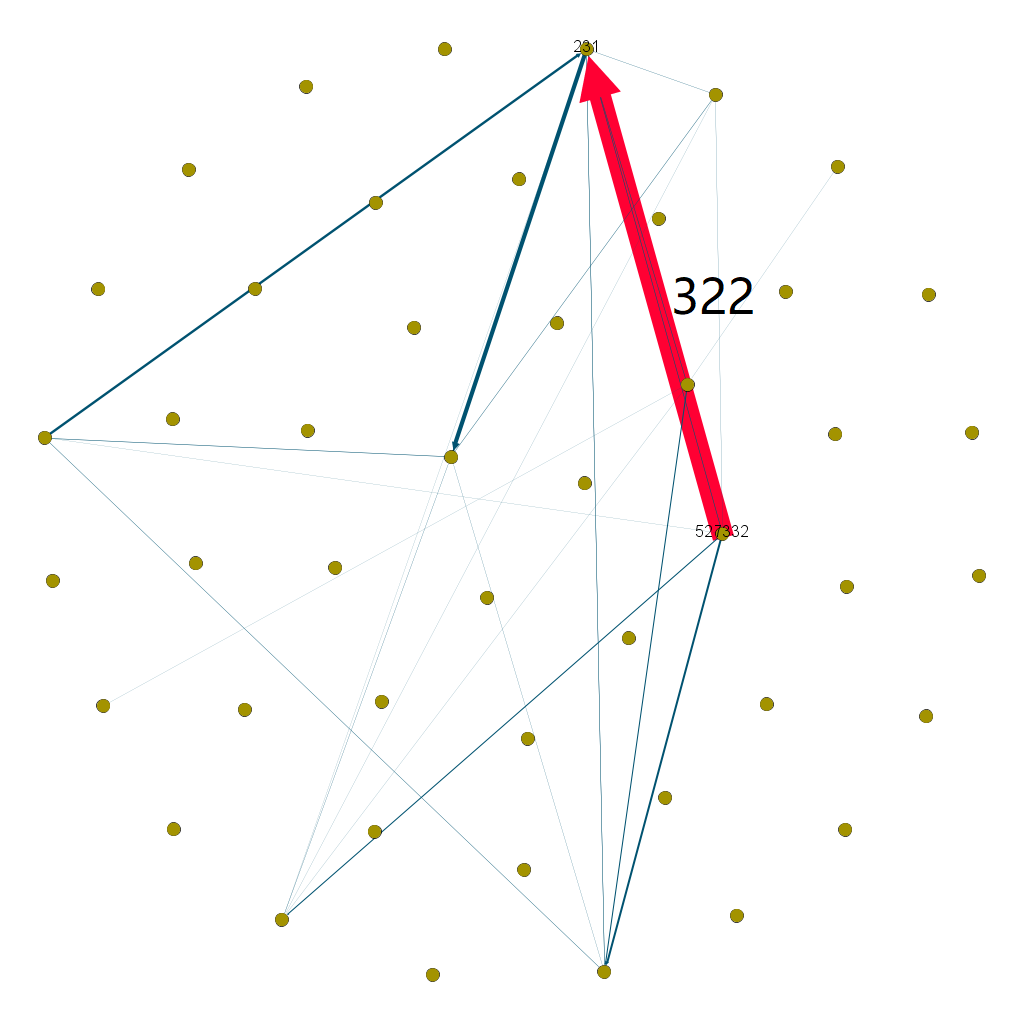}}\hfill
  \subfloat[Bidirection]{%
  \label{Bi-direction}
    \includegraphics[width=.26\textwidth,height=.26\textwidth]{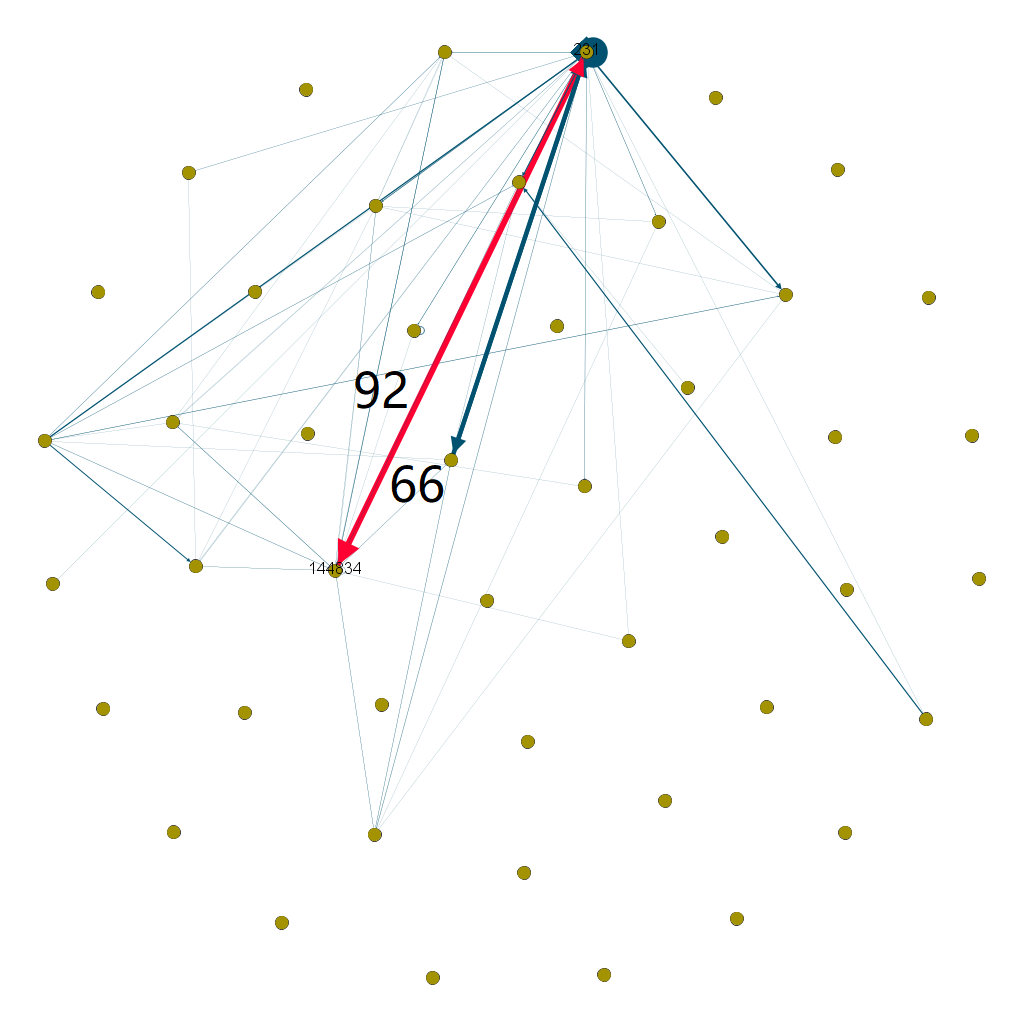}}\

   \subfloat[Triangle]{%
   \label{Triangle}
    \includegraphics[width=.26\textwidth,height=.26\textwidth]{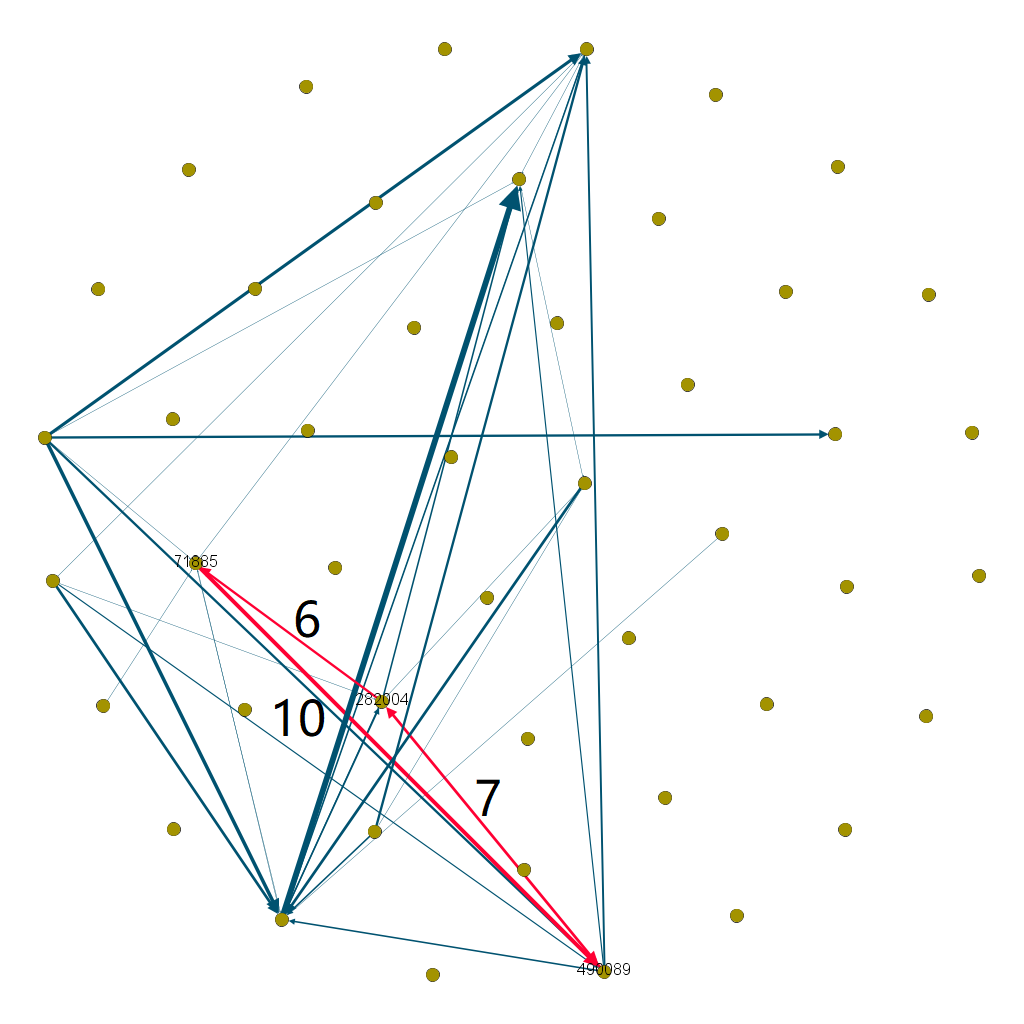}}\hfill
  \subfloat[Polygon]{%
  \label{Polygon}
    \includegraphics[width=.26\textwidth,height=.26\textwidth]{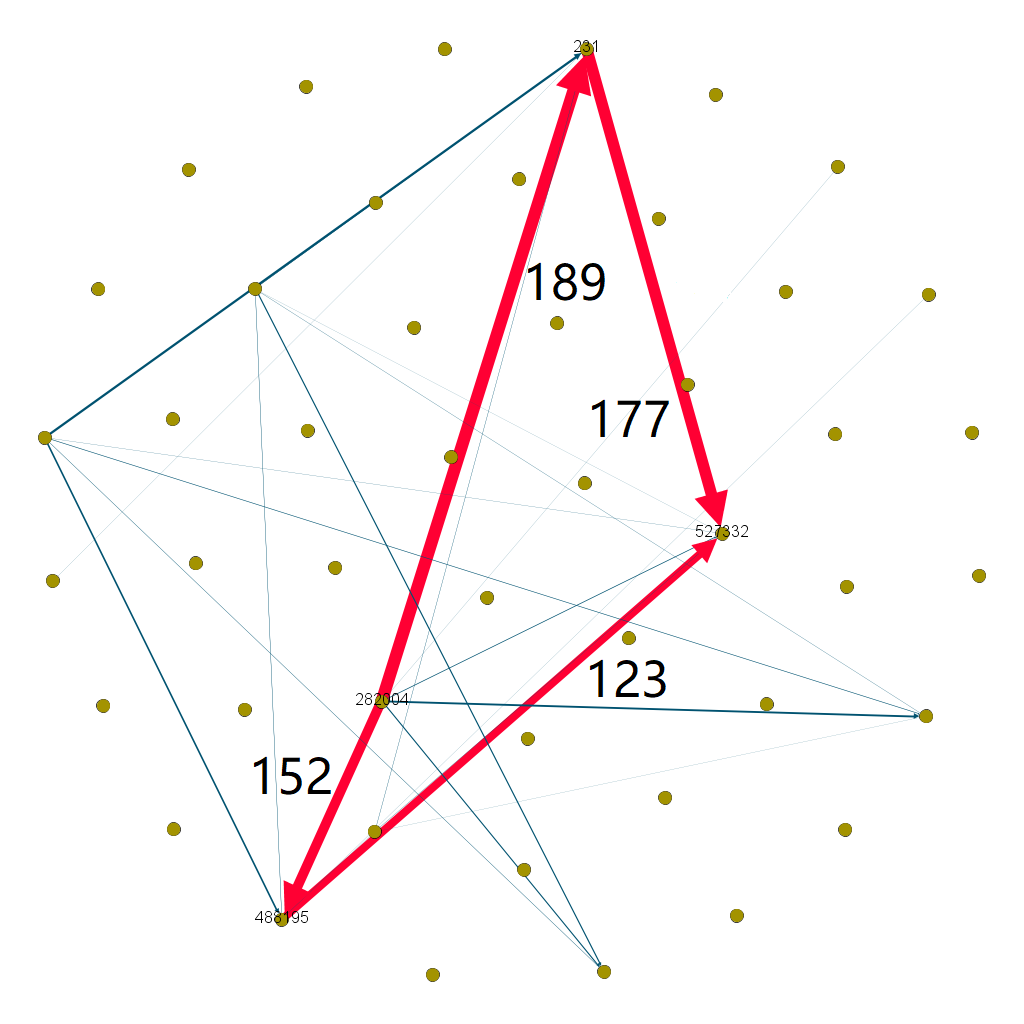}}\hfill
  \subfloat[Star]{%
  \label{Star}
    \includegraphics[width=.26\textwidth,height=.26\textwidth]{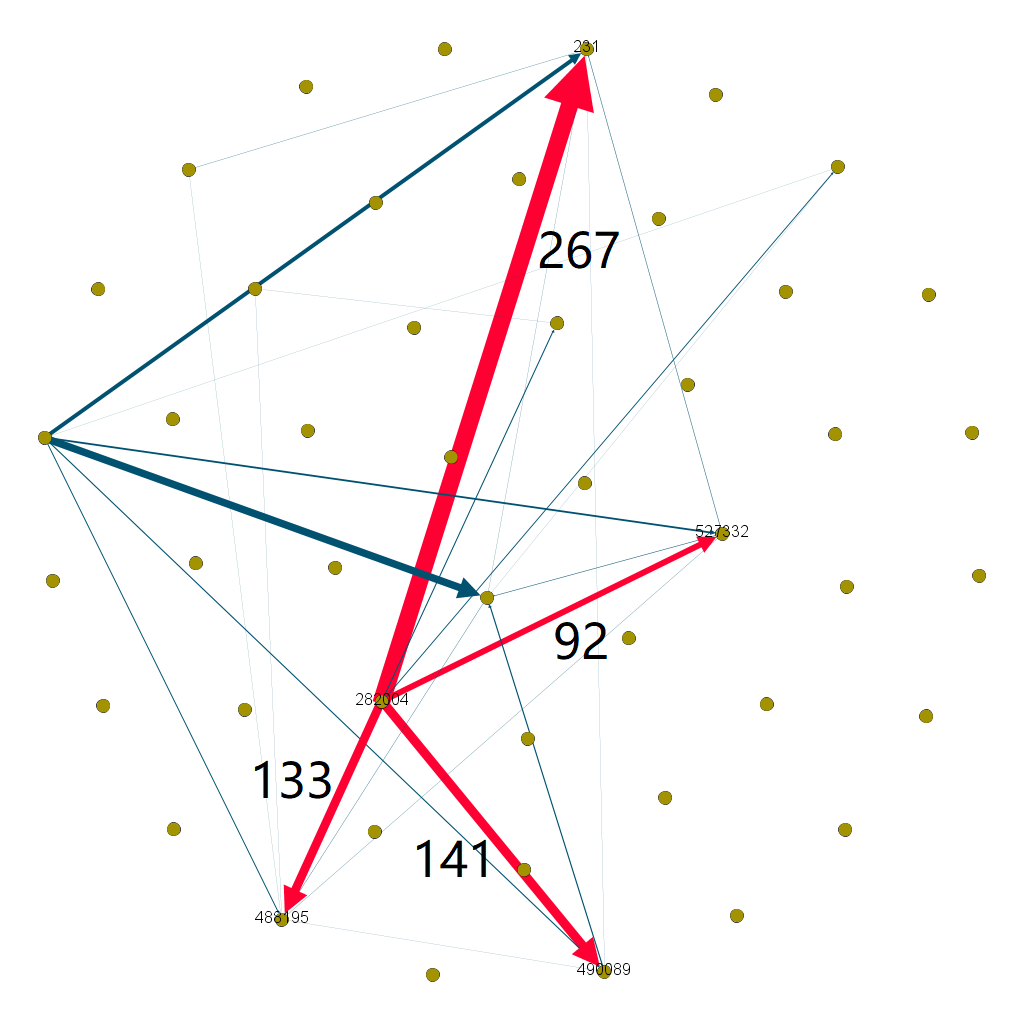}}\

  \caption{Some typical abnormal transaction patterns}\label{tran_pattern}

\end{figure*}

\subsection{Abnormal transaction patterns}
As discussed above, the transactions between abnormal users have a great correlation with the Bitcoin exchange price. A natural question is which edges (i.e, transactions) and thus accounts are the most influential and whether the transactions show certain patterns during the period. To this end, based on the extracted 10 base networks, we further extract the top-10 ranking edges (by the absolute value of weights) in each base networks. We find only 44 distinct edges instead of the 100 maximally possible, which including a total of 28 accounts in EHG. In ELG, 57 edges and 46 accounts were found. We call these \emph{core abnormal accounts}.

To identify special transaction patterns, we draw the daily subgraph of the core abnormal accounts. We find that there are many abnormal transaction patterns (i.e., market manipulation patterns) in the networks. In order to save space, we show only 6 typical patterns in Fig.~\ref{tran_pattern}. These subgraphs are all extracted from ELG on different days. In order to illustrate more clearly, we fix the layout of the graph (i.e., the position of the accounts in each graph is fixed) and denote the special patterns in red. The size of the line denotes the number of transactions between the two accounts. The number at the right-hand side of the directed edge represents the number of transactions between the two accounts. We simply explain the 6 patterns as follows:  
\begin{itemize}
\item \textbf{Self-Loop.} A pattern that an account made transactions with itself. Figure \ref{Self-Loop} shows subgraph on 2013/02/07, the account 231 made 749 transactions with itself. Self-Loop is restricted for normal accounts in any exchanges, as it makes no sense. Thus, a reasonable explanation for the self-loop pattern is that the account may belong to the exchange and may be used to increase daily transaction volume or price manipulation.      
\item \textbf{Unidirection.} The unidirectional pattern indicates more than one transaction from account \emph{A} to \emph{B}. Figure~\ref{Unidirection} shows a unidirectional pattern on 2013/08/15, where account 527332 made 322 sell transactions to account 231. It is possible for an account to sell Bitcoin to another account for more than one times, however, it is almost impossible for two normal accounts to interact with such a large number of times on the same day.       
\item \textbf{Bi-direction.} The bi-directional pattern is a typical market manipulation behavior, especially when the two accounts are controlled by the same user, that two accounts interact with each other many times. Figure \ref{Bi-direction} shows the bi-direction pattern on 2013/04/14 where account 144834 interact with account 231 for more than 150 times. 
\item \textbf{Triangle.} The triangle pattern indicates a triangle-like structure between three accounts. It may contain various forms when considering the direction of the edge. Figure \ref{Triangle} shows a special form of triangle pattern on 2013/10/25. It is special because the accounts form a loop through transactions (account 282004 $\rightarrow $71885 $\rightarrow $490089 $\rightarrow $282004). 
\item \textbf{Polygon.} Polygon pattern is a more complicated transaction pattern where many accounts form a polygon-like \emph{group} with each edge has more than one transactions. Figure \ref{Polygon} shows a quadrangle pattern on 2013/09/19, it seems that account 282004 sends Bitcoin to account 527332 through the ``bridge accounts'' 488195 and 231 for more than two hundred transactions.  

\item \textbf{Star.} A star pattern has a core account that buys or sells Bitcoin to many accounts. Figure~\ref{Star} shows a typical star, where the account 282004 sell Bitcoin to accounts 488195, 490089, 527332 and 231. 

\end{itemize}

Generally speaking, it is not surprising for a transaction network to form a special structure, as transactions are random. However, in our case, it is impossible as each edge represents far more than one transaction in a single day. 
Thus, it seems quite possible that these accounts are controlled by a certain group and these transactions have special purposes.

Based on the results, we summarize the findings as follows:
\begin{itemize}
\item \textbf{Finding 4.} The daily fluctuations of the selected base networks of EHG and ELG have a strong correlation with the Bitcoin exchange price. On the contrary, the daily fluctuation of the base networks of NMG has no correlation with the Bitcoin exchange price. This finding indicates that the behavior of the abnormal accounts' transaction affects the fluctuation of Bitcoin exchange price.     
\item \textbf{Finding 5.} The trend of the Bitcoin exchange price can be captured by the selected base networks of EHG and ELG. It means that the trend of the price can be predicted by transactions between abnormal accounts. 
\item \textbf{Finding 6.} There are many unusual transaction patterns (e.g., self-loop, bi-direction, star) between abnormal accounts. These patterns imply that these accounts are controlled by the same group and are strong evidence of price manipulation.  
\end{itemize} 

\section{Related Work}\label{relatedwork}
Blockchain technology is a new technology, which has many research directions and attracts the interest of researchers from various fields\cite{shaoan,zheng2017overview}. Our research is related to previous work in two areas. The first related area is the study of understanding the big fluctuation of Bitcoin price. As aforementioned, many driving factors of the price are found. Due to all the related data are time series, the most used method in the analysis is time series based model such as vector space model \cite{georgoulausing}, vector error-correction model \cite{ciaian2016economics}, ARDL bounds testing method \cite{bouoiyour2015does}, wavelet analysis \cite{kristoufek2015main}, and vector autoregressive \cite{ciaian2016economics}. 

Another related area is the study of the blockchain data (i.e., the transaction ledger) 
for different topics. Due to the publicly accessible of the blockchain data and users are anonymous in the system, a common topic is to mine the blockchain data to reveal users' privacy \cite{Reidanalysisanonymitybitcoin2013, AndroulakiEvaluatinguserprivacy2013,  AtheyBitcoinpricingadoption2016}. Because of the relatively lawless, blockchain has become an area full of various scams. Thus, mining the blockchain data to detect scams is also a critical topic. Recently, there are many studies on this topic, such as Bitcoin-based scams \cite{VasekTherenofree2015}, the smart contract 
based Ponzi schemes \cite{bartoletti2017dissecting, chen2018detecting}, money laundry \cite{moser2013inquiry}, attacks \cite{Chen2018under}. See \cite{weilisurvey} for a full survey of this topic.

\section{Conclusion and Future Work}\label{conclusion}

We conduct a systematic study to analyze the leaked Mt. Gox transaction data through graph analysis. By comparing the transaction price of the transaction data with the disclosed daily price, many abnormal transactions were identified and were used to divide the accounts into three categories. Based on this classification, we construct three graphs (i.e., EHG, ELG, and NMG) and obtain many findings by analyzing these graphs through various metrics. These findings convinced us that there are many
market manipulation behaviors in the exchange. In order to reveal the relationship between these behaviors and the Bitcoin price, the graphs are reconstructed into daily graph series and reshaped into matrices. Through adopting SVD to the matrices, some very important base networks are identified. By inspecting the base networks, we find that the daily variation of the abnormal base networks closely related to the Bitcoin price and many market manipulation patterns. Based on these findings and considering Bitcoin is dominant in the market, we propose to strengthen supervision in this market. In the future, we will conduct a more thorough study of the data to reveal the extent to which the market is affected and to discuss the changes in the behavior of investors under the extreme fluctuation price.






\section*{Acknowledgment}

The work described in this paper was supported by the National Key Research and Development Program (2016YFB1000101),the National Natural Science Foundation of China (61722214,11801595), the Pearl River S\&T Nova Program of Guangzhou (201710010046) and the Program for Guangdong Introducing Innovative and Entrepreneurial Teams (2016ZT06D211). 
\bibliographystyle{IEEEtran}
\bibliography{NetMtGox}

\end{document}